\documentclass[twocolumn,prb,aps,showpacs,10pt,superscriptaddress]{revtex4-1}
\usepackage{amsfonts}
\usepackage{amsmath}
\usepackage{amssymb}
\usepackage{graphicx}
\usepackage{bm}
\usepackage[usenames,dvipsnames,svgnames,table]{xcolor}
\usepackage{tabularx}

\newcommand{\up}[0]{\uparrow}
\newcommand{\down}[0]{\downarrow}
\newcommand{\mbf}[1]{\mathbf{#1}}
\newcommand{\ddt}[0]{\frac{\partial}{\partial t}}

\renewcommand{\t}[1]{\textrm{#1}}
\renewcommand{\k}[0]{\mbf{k}}

\newcommand{\nn}[0]{\nonumber\\}

\newcommand{\Jsd}[0]{J_{sd}}

\newcommand{\NMn}[0]{N_{Mn}}

\renewcommand{\Re}[0]{\t{Re}}
\renewcommand{\Im}[0]{\t{Im}}
\begin{document}
\title{Carrier-impurity spin transfer dynamics in paramagnetic II-VI 
diluted magnetic semiconductors in the presence of a wave-vector-dependent
magnetic field}
\author{M.~Cygorek}
\affiliation{Theoretische Physik III, Universit{\"a}t Bayreuth, 95440 Bayreuth, Germany}
\author{P.~I.~Tamborenea}
\affiliation{Departamento de F\'isica and IFIBA, FCEN, Universidad de Buenos Aires, Ciudad
Universitaria, Pabell\'on I, 1428 Ciudad de Buenos Aires, Argentina }
\affiliation{Theoretische Physik III, Universit{\"a}t Bayreuth, 95440 Bayreuth, Germany}
\author{V.~M.~Axt}
\affiliation{Theoretische Physik III, Universit{\"a}t Bayreuth, 95440 Bayreuth, Germany}
\begin{abstract}
Quantum kinetic equations of motion for carrier and impurity spins in 
paramagnetic II-VI diluted magnetic semiconductors in a $\k$-dependent
effective magnetic field are derived, where the carrier-impurity correlations
are taken into account. 
In the Markov limit, rates for the electron-impurity spin transfer can be derived
for electron spins parallel and perpendicular to the impurity spins
corresponding to measurable decay rates in Kerr experiments in 
Faraday and Voigt geometry.
Our rigorous microscopic quantum kinetic treatment automatically accounts for
the fact that, in an individual spin flip-flop scattering process, a spin flip of an electron is 
necessarily accompanied by  a flop of an impurity spin in the opposite direction
and the corresponding change of the impurity Zeeman energy influences the
final energy of the electron after the scattering event.
This shift in the electron energies after a spin flip-flop scattering processes, 
which usually has been  overlooked  in the literature, 
turns out to be especially important in the case of extremely diluted magnetic semiconductors in an external 
magnetic field.
As a specific example for a $\k$-dependent effective magnetic field
the effects of a Rashba field on the dynamics of the
carrier-impurity correlations in a Hg$_{1-x-y}$Cd$_y$Mn$_x$Te quantum well
are described. It is found that, although accounting for  
the Rashba interaction in the dynamics of the correlations
leads to a modified $\k$-space dynamics, the 
time evolution of the total carrier spin is not significantly influenced.
Furthermore, a connection between the present theory and the description of 
collective carrier-impurity precession modes is presented.
\end{abstract}
\maketitle
\section{Introduction}
Diluted magnetic semiconductors (DMS) have attracted a great deal of 
interest\cite{Dietl14,DOM,BenCheikh2013,PulsedB_DMS,HyperfineCdMnTe,
SpinWavesPerezCibert11,Perakis_Wang,perakis08,Krenn,Crooker97,Barate10} 
as their highly tunable magnetic properties are ideally suited for adding 
spintronic functionalities to otherwise well-established 
semiconductor technologies\cite{Zutic,spintronics_dietl,spintronics_ohno}.
Particularly promising for future technological applications is 
the fact that some DMS, such as Ga$_{1-x}$Mn$_x$As, exhibit a ferromagnetic
phase\cite{DOM,CdMnTeFerroMag97}. The convenient optical properties 
also allow, e.g., for the optical switching of the 
magnetization\cite{Perakis2012} in Ga$_{1-x}$Mn$_x$As.
While a comprehensive unified theoretical description of the magnetism 
in DMS is still missing, it is generally accepted that a
carrier-mediated impurity-impurity spin interaction plays a key 
role\cite{JungwirthRevModPhys,Dietl14}.
Thus, it is crucial to understand the spin physics not only of
the magnetic impurities, but also of the carriers as well as the 
details of the spin transfer between carriers and impurities.

Experimentally, the carrier spins in DMS are often investigated 
optically using time-resolved magneto-optical Kerr effect (MOKE) 
measurements\cite{Patz15,Crooker97,BenCheikh2013}, a pump-probe 
technique that makes it possible to extract the carrier spin dynamics
with a temporal resolution of $\sim$ 100 fs. The experimentally obtained 
carrier spin dephasing and relaxation rates, which also include the effects
of the spin transfer between carriers and impurities,
can then be used as an input for, e.g., 
the theoretical description of spin wave excitations in ferromagnetic
DMS\cite{perakis08}. 

However, a quantitative theoretical explanation for the values 
of the carrier spin relaxation rates measured in MOKE experiments, 
even in the simplest possible case of 
conduction band electrons in an intrinsic II-VI DMS, has yet to be 
found. 
%
For example, even such basic quantities as 
the magnetic field dependence of the spin transfer rate between the 
carrier and impurity systems in paramagnetic DMS is still not satisfactorily 
explained\cite{BenCheikh2013}.
This is, on the one hand, due to the large number of factors that 
simultanously play a role in DMS, like the
spin-dependent $s$-$d$ interaction between magnetic 
impurities and carriers,
spin dephasing due to spin-orbit coupling mechanisms\cite{Rashba,Dresselhaus,
Ungar15}, carrier-carrier interaction\cite{Wu09} and disorder
effects\cite{Roennburg}. 
On the other hand, even the typically dominant $s$-$d$ interaction is
usually treated only on the level of the mean-field 
approximation\cite{FurdynaReview,Koenig,Morandi10_2}, 
neglecting the effects of carrier-impurity correlations, which can be 
important\cite{perakis08,Morandi10,FreqRenorm}. 
The spin transfer between carriers and impurities is
commonly described by rate equations where the rates are calculated 
using Fermi's golden rule\cite{Koenig,TsitsishviliExcDMS,
TsitsishviliSpinConsHoles,Morandi10_2,KossutRate3D}.

One problem of this approach
is that it is {\it a priori} not clear under which circumstances the
perturbative scheme, which is implicit in the derivation of Fermi's golden
rule, is applicable. For example, at the band edge, where the band energies, 
described by the Hamiltonian $H_0$ of an undoped semiconductor, are
negligible, the $s$-$d$ interaction 
cannot be thought of as a small
perturbation to $H_0$. A second deficiency of the golden-rule treatment is
that it gives, by construction, only the transition rate between energy
eigenstates of the system. However, optical orientation also allows for
an injection of carrier spins perpendicular to an external magnetic field
(Voigt geometry) or the impurity magnetization, 
respectively\cite{Crooker97}, which corresponds to the excitation 
of superpositions of energy eigenstates.
Thus, the relaxation rate of the transverse carrier spin component is
not provided by Fermi's golden rule. 

A more elaborate treatment of the $s$-$d$ exchange interaction, which 
is also capable of deriving a rate for the spin transfer of the perpendicular
electron spin component, was given {by \it Semenov} in a study based on a
projection operator method\cite{Semenov}.
Another notable approach to the spin dynamics in DMS has been provided by
the group of {\it Wu}\cite{Wu09}, which has developed the kinetic spin 
Bloch equations (KSBEs) that account not only for rates for the spin transfer
due to the $s$-$d$ exchange interaction, but also for a number of other effects,
such as carrier-carrier and carrier-phonon interaction.

In the present article, we describe the electron spin dynamics
in the conduction band,
where we focus on 
paramagnetic intrinsic II-VI DMS.
We work with a quantum kinetic theory starting from the
$s$-$d$ exchange Hamiltonian $H_{sd}$, where a correlation expansion 
scheme was used to formulate equations of motion for the carrier and 
impurity density matrices as well as the carrier-impurity correlation 
functions\cite{Thurn:12}.
This approach allows a 
non-perturbative description of far-from-equilibrium situations. 
The golden-rule rate equations can be deduced from the quantum kinetic 
theory as a Markovian limit\cite{Thurn:13_1,Thurn:13_2}. In the same limit,
also the rates for the carrier spin component perpendicular to the impurity
magnetization can be derived\cite{Cygorek:14_1}. 
Furthermore, the applicability of the Markovian
limit and therewith the golden-rule rate equations can be checked by
direct comparison of the full quantum kinetic theory with its Markovian 
limit\cite{Cygorek:14_1}.  It was found that for an agreement between 
the quantum kinetic and the Markovian predictions, it is essential to
account for a precession-like motion of the carrier-impurity 
correlations. Therefore, effective equations which capture the essential 
features of the full quantum kinetic equations that also include 
the correlation dynamics, were called {\it precession of electron spins and
correlations} (PESC) equations\cite{PESC}.

For vanishing external magnetic field and impurity magnetization, 
all of the above theories contain the same rate equations that can also be
found with Fermi's golden rule as a special case.
In contrast, in the presence of an external magnetic field which leads to
a finite impurity magnetization in the equilibrium of a paramagnetic DMS,
the predictions of the different theories deviate from each other.
In order to compare these theories, 
we extend the quantum kinetic theory of Ref.~\onlinecite{Thurn:12} to take into 
account the Zeeman interaction of carriers and impurities in a magnetic 
field. 

We also allow for a $\k$-dependence of
an effective magnetic field, which makes it possible to discuss the effects
of Dresselhaus\cite{Dresselhaus} or Rashba\cite{Rashba}
spin-orbit coupling or a $\k$-dependent g-factor 
on the spin dynamics in DMS. 
In contrast to previous treatments\cite{Ungar15} where the PESC equations
were extended by adding a $\k$-dependent precession term to the 
time evolution of the carrier spins, in the present article the $\k$-dependent
effective magnetic field is incorporated on a microscopic quantum kinetic 
level which also leads to a modification of the equations of motion
for the carrier-impurity correlations. Another point of view is that, 
while the approach of Ref.~\onlinecite{Ungar15} accounts for the 
$\k$-dependent field \textit{between} 
carrier-impurity spin-flip scattering events,
in the present theory the effective magnetic field also acts \textit{during}
the spin-flip scattering.
Formally this situation is similar to that of, e.g., 
the intracollisional field effect\cite{RossiICFE}, where 
the effects of an external field that acts during a scattering event
(phonon-emission in the case of Ref.~\onlinecite{RossiICFE}) can indeed change
the optical and transport properties qualitatively. 

Furthermore, here, we account for the fact that
the impurity spin is a $z$-dependent 
(growth direction of the quantum well) dynamical variable which can change over
time. This connects the present theory to the description of 
collective carrier-impurity precession 
modes\cite{ColMod2,ColMod_Teran,Barate10}.

The article is outlined as follows:
First, we derive the Markov limit of quantum kinetic equations 
accounting for the $s$-$d$ interaction, a possibly $\k$-dependent 
effective magnetic field and the $z$-dependence of the carrier
envelope function.
Then, we present results for the magnetic field dependence of the
carrier-impurity spin transfer rates and compare it with the results 
predicted by several other theories.
Next, we answer the question to what extent spin-orbit couplings 
that lead to a $\k$-dependent effective magnetic field influence the
spin transfer dynamics, in particular with respect to the 
dynamics of the carrier-impurity correlations.
Finally, we show how the theory of the present paper can be related to
the theory employed in the discussion of collective carrier-impurity
precession modes\cite{ColMod2}.

\section{Theory}
\subsection{DMS Hamiltonian}
The Hamiltonian for electrons and impurities in DMS can be modelled by
\begin{align}
&H=H_0+H_Z^e+H_Z^{Mn}+H_{sd},
\end{align}
where $H_0$ describes the conduction band of a semiconductor crystal
and can be written as 
\begin{align}
&H_0=\sum_{\k}\sum_{\sigma}
\hbar\omega_{\k} c^{\dagger}_{\k\sigma}c_{\k\sigma}+
\sum_{\k}\sum_{\sigma,\sigma'}
\hbar\boldsymbol\Omega_{\k}\cdot\mbf s_{\sigma\sigma'}
c^{\dagger}_{\k\sigma}c_{\k\sigma'}.
\end{align}
$c^{\dagger}_{\k\sigma}$ and $c_{\k\sigma}$ are the creation and annihilation
operators for electrons with wave vector $\k$ in the spin subband 
$\sigma\in\{\up,\down\}$.
$\omega_{\k}$ describes the diagonal, i.e., the spin independent, part of
$H_0$ while $\boldsymbol\Omega_\k$ is the $\k$-dependent effective magnetic
field, e.g., due to spin-orbit interactions. The electron spin matrix vector
$\mbf s_{\sigma\sigma'}=\frac 12\boldsymbol\sigma_{\sigma\sigma'}$ 
is proportional to the vector of Pauli matrices 
$\boldsymbol\sigma_{\sigma\sigma'}$\footnote{
Here, we use the convention that the factor $\hbar$ which appears in the 
spin matrices in the SI system is instead included 
in $\mu_B$ and $\Jsd$, respectively.
}.

$H^e_Z$ and $H^{Mn}_Z$ are the Zeeman energies for carriers and impurities,
respectively:
\begin{align}
&H_Z^e=\sum_{\k\sigma\sigma'} g_e(\k) \mu_B\mbf B\cdot \mbf s_{\sigma\sigma'}
c^\dagger_{\k\sigma}c_{\k\sigma'}, \\
&H_Z^{Mn}=\sum_{I nn'}g_{Mn}\mu_B\mbf B\cdot \mbf S_{nn'} \hat{P}^I_{nn'},
\end{align}
where $g_e$ and $g_{Mn}$ are the electron and impurity g-factors and
$\mbf B$ is the externally applied magnetic field. In general, 
$g_e$ may depend on the electron wave vector which, e.g., gives rise to
the imhomogeneous-g-factor spin dephasing mechanism\cite{Margulis,Ogg}
which is essential for the description of the 
magnetic field dependence of the spin decay time in nonmagnetic
semiconductors\cite{Bronold02}.
$\mbf S_{nn'}$ are the spin matrices for the impurities with, in 
the case of Manganese, S=$\frac 52$, so that $n,n'\in\{-\frac 52,-\frac 32,
\dots,\frac 52\}$. The impurity spin is described by the
operator $\hat{P}^I_{nn'}=|I,n\rangle\langle I,n'|$ 
where $|I,n\rangle$ is the $n$-th spin state of the $I$-th impurity ion.

The most important part of the Hamiltonian for the spin dynamics in DMS 
is the $s$-$d$ exchange interaction which, in real space, has the form:
\begin{align}
&H_{sd}=\Jsd\sum_{\substack{I,n,n',\\ i,\sigma,\sigma'}}
\big(\mbf S_{nn'} \hat{P}^I_{nn'}\big)\cdot \mbf s_{\sigma\sigma'} 
\psi^\dagger_\sigma(\mbf r_i) \psi_{\sigma'}(\mbf r_i)
\delta(\mbf R_I -\mbf r_i),
\end{align}
where $\mbf R_I$ and $\mbf r_i$ are the position vectors of the $I$-th
impurity and  $i$-th electron and $\psi^\dagger_\sigma(\mbf r_i)$
as well as $ \psi_{\sigma}(\mbf r_i)$ are the corresponding real-space
field operators for the electrons.
Since most experiments on DMS are performed on two-dimensional structures,
we choose a single-particle basis comprised of product states of a 
$z$-dependent envelope, where $z$ is defined
to point along the growth direction, and an in-plane part described by
plane waves. 
When restricting to the lowest confined state of the 
envelope function $\psi(z)$, we can formulate the effective $s$-$d$ Hamiltonian
for the in-plane part as:

\begin{align}
&H_{sd}=
\frac{\Jsd}V d\sum_I |\psi(Z_I)|^2 \mbf S_{nn'}\cdot \mbf s_{\sigma\sigma'}
c^\dagger_{\k\sigma}c_{\k'\sigma'} \hat{P}^I_{nn'}e^{i(\k'-\k)\mbf R^\|_I},
\end{align}
where $V$ is the volume of the sample, $d$ is the quantum well width, 
$Z_I$ is the z-component 
of the $I$-th impurity position vector and $\mbf R^\|_I$ is the in-plane part
of the position vector of the $I$-th impurity.
Assuming infinitely high barriers, the envelope is given by
\begin{align}
&\psi(z)=\sqrt{\frac{2}d}\cos\Big(\frac\pi d z\Big),
\end{align}
for $z\in [-\frac d2;\frac d2]$ and zero otherwise.
Thus, due to the factor $|\psi(Z_I)|^2$, magnetic impurities
at the border of the quantum well couple much more weakly to the electrons than
impurities at the center of the well.

\subsection{Quantum Kinetic Equations of Motion}
In Ref.~\onlinecite{Thurn:12}, a set of quantum kinetic equations of motion
based on a correlation expansion scheme has been developed for the
carrier and impurity density matrix as well as the carrier-impurity 
correlations in the case 
of zero external and effective-spin-orbit magnetic fields.
In the present article, we additionally consider an in general wave-vector 
dependent effective magnetic field for the carriers and the Zeeman energy term 
for the magnetic impurities to the Hamiltonian.
Since all of the terms that are added are effective single-particle
contributions, they do not lead to a build-up of a new hierarchy of 
correlations, but only connect the density matrices and the correlations
with themselves. 
Therefore, the trucation scheme and the factorization of higher 
correlations layed out in Ref.~\onlinecite{Thurn:12} can 
still be applied when the aforementioned 
additional Hamiltonians are accounted for.
If an on average homogeneous distribution of magnetic impurities in 
the quantum-well plane 
is assumed,
equations of motion can be formulated for the dynamical 
variables\cite{Cygorek:14_1}
\begin{subequations}
\begin{align}
C_{\sigma_1\k_1}^{\sigma_2}=&\langle c^\dagger_{\k_1\sigma_1}
c_{\k_1\sigma_2}\rangle, \\
M_{n_1}^{n_2}(z)=&\frac d{\NMn}
\sum_I\delta(z-Z_I)\langle \hat{P}^I_{n_1n_2} \rangle, \\
Q_{\sigma_1n_1\k_1}^{\sigma_2n_2\k_2}(z)=&
\frac{V}{\NMn}d\sum_I\delta(z-Z_I)\times\nn&
\langle c^\dagger_{\k_1\sigma_1}c_{\k_2\sigma_2}
\hat{P}^I_{n_1n_2}e^{i(\k_2-\k_1)\mbf R_I^\|}\rangle,
\end{align}
\label{eq:olddynvar}
\end{subequations}
where $C_{\sigma_1\k_1}^{\sigma_2}$ and $M_{n_1}^{n_2}(z)$  are the 
electron and impurity density matrices and $Q_{l_1n_1\k_1}^{l_2n_2\k_2}(z)$
(for $\k_1\neq\k_2$) 
represent the carrier-impurity correlations, where the 
mean-field part has been subtracted. 
$\NMn$ is the number of impurity ions in the DMS.

Instead of the density matrices, also the average carrier $\mbf s_{\k_1}$ 
and impurity spins $\langle \mbf S(z)\rangle$ as well as the electron
occupations $n_{\k_1}$ can be used as dynamical variables\cite{Cygorek:14_1}
which helps to understand the dynamics of the physical variables and
simplifies the equations of motion.
\begin{subequations}
\begin{align}
&\langle \mbf S(z)\rangle=\sum_{nn'} \mbf S_{nn'}M_n^{n'}(z),\\
&n_{\k_1}=\sum_{\sigma}C_{\sigma\k_1}^{\sigma},\\
&\mbf s_{\k_1}=\sum_{\sigma_1\sigma_2}\mbf s_{\sigma_1\sigma_2}
C_{\sigma_1\k_1}^{\sigma_2},\\
&Q^{\alpha \k_2}_{j\k_1}:= \sum_{\stackrel{\sigma_1\sigma_2}{n_1n_2}}
S^j_{n_1n_2}s^\alpha_{\sigma_1\sigma_2} 
Q^{\sigma_2n_2\k_2}_{\sigma_1n_1\k_1}.
\end{align}
\label{eq:newdynvar}
\end{subequations}

From now on, we will use the convention that $\sigma$-indices
describe spin-up and spin-down subbands, 
$n$-indices enumerate the impurity states,
while all other Latin indices represent three-dimensional geometric 
directions, e.g., $j\in\{1,2,3\}$, and Greek indices range from $0$ to $3$,
where the $0$ describes occupations. In this notation, the zeroth
spin matrix is defined to be the $2\!\times\!2$
identity matrix $s^0_{\sigma_1\sigma_2}=\delta_{\sigma_1\sigma_2}$. 
Furthermore, we adopt the Einstein notation, so
that when the same index appears twice, a summation is implied. Sub- and 
superscripts are used, e.g., to distinguish the carrier and impurity degrees 
of freedom of the correlations, and do not represent a
covariant formulation.
Sums over $\k$ vectors, on the other hand, will be stated explicitly and
no sum is implied, if an index $\k_i$ appears twice in an expression.

In this notation, the equations of motion of
Ref.~\onlinecite{Cygorek:14_1,PESC}, extended by terms 
due to the $\k$-dependent effective magnetic field and the impurity 
and carrier Zeeman energies, are:
\begin{widetext}
\begin{subequations}
\begin{align}
&\ddt \langle S^l(z)\rangle=
\Big(\boldsymbol\omega_{Mn}(z)\times \langle \mbf S(z)\rangle\Big)_l
-\frac{\Jsd|\psi(z)|^2d}{\hbar V^2} 
\sum_{\k\k'}\epsilon_{ijl}\Re\big(Q_{i\k}^{j\k'}(z)\big),\\
&\ddt n_{\k_1}= \int\limits_{-\frac d2}^{\frac d2}dz\;
\frac{\Jsd|\psi(z)|^2\NMn}{\hbar V^2} \sum_\k
2\Im\big(Q_{i\k_1}^{i\k}(z)\big),\\
&\ddt s^l_{\k_1}=\Big(\boldsymbol\Omega'_{\k_1}\times \mbf s_{\k_1}\Big)_l
+\int\limits_{-\frac d2}^{\frac d2}dz\; \frac{\Jsd|\psi(z)|^2\NMn}{\hbar V^2} 
\sum_\k \Im\bigg[\frac 12 Q_{l\k_1}^{0\k}(z)+
i \epsilon_{ijl} Q_{i\k_1}^{j\k}(z)\bigg], \\
&\ddt Q_{l\k_1}^{\alpha\k_2}(z)=
-i(\omega_{\k_2}-\omega_{\k_1})Q_{l\k_1}^{\alpha\k_2}(z)+
\big(A_{\k_1}+A_{\k_2}^*\big)_{\alpha\gamma}Q_{l\k_1}^{\gamma\k_2}(z)
+\epsilon_{ijl} 
\omega_{Mn}^i(z) Q_{j\k_1}^{\alpha\k_2}(z)
+b_{l\k_1}^{\alpha\k_2}(z)
+c_{l\k_1}^{\alpha\k_2}(z)
,\\
&b_{l\k_1}^{\alpha\k_2}(z)=\frac i\hbar \Jsd d|\psi(z)|^2\Big[
\langle S^iS^l (z)\rangle \langle s^i s^\alpha\rangle_{\k_2}^{\k_1}
-\langle S^l S^i (z)\rangle \langle s^\alpha s^i\rangle_{\k_1}^{\k_2}\Big],
\end{align}
\label{eq:eom}
\end{subequations}
\end{widetext}
where the mean-field precession frequencies for impurities and carriers
are defined as
\begin{subequations}
\begin{align}
&\boldsymbol\omega_{Mn}(z):=\frac{g_{Mn}\mu_B}\hbar \mbf B+
\frac{\Jsd|\psi(z)|^2 d}{\hbar V}\sum_\k \mbf s_{\k}, \\
&\boldsymbol\Omega'_{\k}:=\boldsymbol\Omega_{\k}+\boldsymbol\omega_e(\k)\\
&\boldsymbol\omega_{e}(\k):=\frac{g_{e}(\k)\mu_B}\hbar \mbf B+
\int\limits_{-\frac d2}^{\frac d2}dz\; \frac{\Jsd|\psi(z)|^2 \NMn}{\hbar V}
\langle \mbf S(z)\rangle,
\label{eq:def_om_e}
\end{align}
The $\k$-dependent precession-like movement of the electron degree of 
freedom of the correlations is described by the 4$\times$4 matrix 
\begin{align}
A_{\k_1}:=\left(\begin{array}{cc}
0& (i\boldsymbol\Omega'_{\k_1})^T  \\
(\frac i4\boldsymbol\Omega'_{\k_1}) & 
\frac 12[\boldsymbol{\Omega'_{\k_1}}]_\times 
\end{array}\right),
\label{eq:defA}
\end{align}
\end{subequations}
where $[\boldsymbol{\Omega'_{\k_1}}]_\times$ is the 3$\times$3 
cross-product matrix
with $[\boldsymbol{\Omega'_{\k_1}}]_\times \mbf v=\boldsymbol{\Omega'_{\k_1}}
\times \mbf v$.

The source terms $b_{l\k_1}^{\alpha\k_2}(z)$ involve
electron variables $n_{\k}$ and $\mbf s_{\k}$ in the form:
\begin{subequations}
\begin{align}
&\langle s^i s^j\rangle_{\k_1}^{\k_2}\!:=
\delta_{ij}\Big[\frac 14\big(1\!-\!\frac{n_{\k_2}}2\big)n_{\k_1}
+\frac 12\mbf s_{\k_1}\!\!\cdot\!\mbf s_{\k_2}\Big] -\frac 12
s^i_{\k_1}s^j_{\k_2}+\nn&-\frac 12s^j_{\k_1} s^i_{\k_2}
+\frac i2
\epsilon_{ijl}\Big[\big(1-\frac{n_{\k_2}}2\big)
s^l_{\k_1}+\frac{n_{\k_1}}2 s^l_{\k_2}\Big],
\end{align}
and
\begin{align}
&\langle s^i s^0\rangle_{\k_2}^{\k_1}:=
\big(1-\frac{n_{\k_1}}2\big) s^i_{\k_2}
-\frac{n_{\k_2}}2s^i_{\k_1}
-
i\epsilon_{ijl} s^j_{\k_1} 
s^l_{\k_2},\\
&\langle s^0 s^i \rangle_{\k_1}^{\k_2}:=
\big(1-\frac{n_{\k_2}}2\big) s^i_{\k_1}
-\frac{n_{\k_1}}2s^i_{\k_2}
-
i\epsilon_{ijl}s^j_{\k_1}
 s^l_{\k_2}.
\end{align}
\end{subequations}
Also, $b_{l\k_1}^{\alpha\k_2}(z)$ contains second moments of the 
impurity variables:

\begin{align}
&\langle S^i S^j(z)\rangle=\langle {S^\perp}^2(z)\rangle
\delta_{ij}+\langle {S^\|}^2(z)-{S^\perp}^2(z)\rangle\times\nn&
\frac{\langle S^i(z)\rangle\langle S^j(z)\rangle}{\langle \mbf S(z)\rangle^2}
+\frac i2
\epsilon_{ijl} \langle S^l(z)\rangle,
\label{eq:Ssec}
\end{align}
where $S^\|:=\frac{\mbf S \cdot\langle\mbf S\rangle}{\langle \mbf S\rangle^2}$ 
is the spin operator projected onto the direction of the average impurity spin
and $\langle {S^\perp}^2\rangle=\frac 12\langle  S^2-{S^\|}^2\rangle$ is the
perpendicular second moment, with $\langle S^2\rangle=\frac{S(S+1)}4=
\frac{35}4$ for a spin-$\frac 52$ system.

By going over from the density matrices in 
Eqs.~(\ref{eq:olddynvar}) as dynamical variables to the 
variables defined in Eqs.~(\ref{eq:newdynvar}), one ends up with a 
set of equations that is not closed. Thus, some approximations have 
to be employed to evaluate the right-hand side of Eqs.~(\ref{eq:eom}): 
First of all, it is necessary to evaluate the 
second moments of the impurity magnetization, for which the equations
of motion can in principle be calculated, but they involve even higher
moments. We reduce the complexity of the equations by calculating a
quasi-thermal impurity density matrix in each time step, which is
consistent with the average spin $\langle \mbf S(z)\rangle$.
Furthermore, the source terms $c_{l\k_1}^{\alpha\k_2}(z)$\footnote{
The source terms $c_{l\k_1}^{\alpha\k_2}$ are given by
$c_{l\k_1}^{\alpha\k_2}:=\sum_{\sigma_1\sigma_2 n_1n_2}
S^l_{n_1n_2} s^\alpha_{\sigma_1\sigma_2}
{b_{\sigma_1n_1\k_1}^{\sigma_2n_2\k_2}}^{III}$ with
${b_{\sigma_1n_1\k_1}^{\sigma_2n_2\k_2}}^{III}$ being defined in 
Ref.~\onlinecite{Cygorek:14_1}.
}
contain degrees of freedom of the original 
correlation functions $Q_{\sigma_1 n_1\k_1}^{\sigma_2 n_2\k_2}$ that are
not expressible in terms of $Q_{l\k_1}^{\alpha\k_2}$. However, the terms 
$c_{l\k_1}^{\alpha\k_2}(z)$ were shown to be irrelevant in numerical 
calculations in the situation described in Ref.~\onlinecite{PESC}.
Since these terms are proportional to some correlation functions
$Q_{\sigma n \k}^{\sigma' n' \k'}$, they mainly renormalize the 
frequencies with which the correlations oscillate. As will be seen later, 
the values of these frequencies determine the difference in kinetic energies
of the initial and final states of carriers scattered due to the $s$-$d$ 
interaction. On the other hand it will be shown that neglecting the 
terms $c_{l\k_1}^{\alpha\k_2}(z)$ leads to equations that conserve the
mean-field energies of the carriers, so that the role of these terms is
mainly to ensure energy conservation including the carrier-impurity
correlation energy. However, this correlation energy is typically
of the order of a few $\mu$eV\cite{FreqRenorm}, so that it is typically
a good approximation to neglect the source terms $c_{l\k_1}^{\alpha\k_2}(z)$,
as we will henceforth do.

With these approximations, it seems straightforward to solve the coupled 
system of ordinary differential equations~(\ref{eq:eom}) numerically.
However, this task is very challenging, since the correlations are indexed by 
two $\k$-vectors, where each one is an element of a two-dimensional continuum 
in the case of a quantum well. The problem therefore has the complexity
$\mathcal{O}(N_k^4 N_z N_t)$, where $N_k$, $N_z$ and $N_t$ are the numbers
of discretization points of the k-space (linear dimension), the growth-direction
in real space and the time, respectively. 
Our strategy to make the calculation tractable follows
Ref.~\onlinecite{PESC}: 
The computation time can be strongly reduced, 
if the correlations are eliminated and only their effects on the electron and 
impurity variables are kept. This can be achieved by formally integrating the
equations of motion for the correlations at the cost of introducing a memory integral.
This memory integral can in turn be eliminated by a short-memory or Markov approximation,
which is established in the next section.

\subsection{Derivation and Applicability of the Markov limit}
Before we discuss the Markov limit of the correlations including the
precession-like movement of the correlations,
we briefly recapitulate the standard way\cite{RossiKuhn02,Cygorek:14_1}
of deriving the Markov limit
of quantum kinetic equations in the simplest possible situation
with $\boldsymbol\Omega'_{\k}=0$ and $\boldsymbol\omega_{Mn}(z)=0$. 
There, the equation of motion (\ref{eq:eom}d) for the correlations becomes
\begin{align}
\ddt Q_{l\k_1}^{\alpha\k_2}=
-i(\omega_{\k_2}-\omega_{\k_1})Q_{l\k_1}^{\alpha\k_2}
+b_{l\k_1}^{\alpha\k_2}.
\end{align}
If the source term $b_{l\k_1}^{\alpha\k_2}$ is regarded as a time-dependent
inhomogeneity, one can first solve the homogeneous part of the equation and
take the inhomogeneity into account by a variation of constants, which yields:
\begin{align}
Q_{l\k_1}^{\alpha\k_2}(t)=&e^{-i(\omega_{\k_2}-\omega_{\k_1})t}
\Big[Q_{l\k_1}^{\alpha\k_2}(t_0)+ \nn&+
\int\limits_{t_0}^t dt'\,
e^{i(\omega_{\k_2}-\omega_{\k_1})t'} b_{l\k_1}^{\alpha\k_2}(t')\Big].
\end{align}
We assume that the carriers stem exclusively form optical excitation and
therefore also the correlations are zero before the laser pulse is applied.
Therefore,  $Q_{l\k_1}^{\alpha\k_2}(t_0)=0$ for $t_0\to -\infty$.
The correlations act back on the carrier and impurity variables only
via sums over correlations with respect to at least one $\k$-index.
Thus, we consider, e.g.,
\begin{align}
\sum_{\k_2}Q_{l\k_1}^{\alpha\k_2}(t)=
\int\limits_{0}^{\omega_{BZ}}\!d\omega\, D(\omega)\!\!
\int\limits_{-\infty}^t\! dt'\ 
e^{i(\omega-\omega_{\k_1})(t'-t)} b_{l\k_1}^{\alpha\k(\omega)}(t'),
\label{eq:sumkQ}
\end{align}
with the quasi-continuous limit
\begin{align}
\sum_{\k} \dots \to \int\limits_{BZ} dk\, D(k) \dots=
\int\limits_0^{\omega_{BZ}}d\omega\, D(\omega) \dots,
\end{align}
where $\hbar\omega$ are the spin-independent single-particle energies
of $H_0$ and $\hbar\omega_{BZ}$ is a cut-off energy corresponding to 
the upper end of the conduction band. Although this expression is valid
also for non-parabolic band structures, we simplify the discussion by
first assuming an effective mass approximation in two dimensions, 
so that $D^{2D}:=D(\omega)=\frac{Am^*}{2\pi\hbar}$ is constant.

Now, the Markov or short-memory approximation can be applied to 
Eq.~(\ref{eq:sumkQ}) as follows:
Assuming that, because of the $\k$-sum, the effects of the correlations 
on the carrier and impurity dynamics dephase fast for not too small 
values of $t'-t$ in the integral kernel, the largest contribution 
of the integrals stems from source terms $b_{l\k_1}^{\alpha\k(\omega)}(t')$
with $t'\approx t$. Then, Eq.~(\ref{eq:sumkQ}) can be approximated by
\begin{align}
&\sum_{\k_2}Q_{l\k_1}^{\alpha\k_2}(t)\approx 
D^{2D}\int\limits_{0}^{\omega_{BZ}}\!d\omega\,
b_{l\k_1}^{\alpha\k(\omega)}(t) \!
\int\limits_{-\infty}^t\! dt'\, e^{i(\omega-\omega_{\k_1})(t'-t)} 
\label{eq:sumQmarkov}
\end{align}
Using the Sokhotski-Plemelj formula
\begin{align}
&\int\limits_{-\infty}^0\! dt' e^{i x t'}=
\pi\Big(\delta(x)-\mathcal{P}\frac{i}{\pi x}\Big)=:\pi\bar{\delta}(x),
\label{eq:Plemelj}
\end{align}
where $\mathcal{P}$ denotes the Cauchy principal value, 
allows to express the correlations solely in terms of carrier and 
impurity variables evaluated at $t'=t$. For the real part of $\bar{\delta}$,
the $\k$-sum reduces to an integration over a single energy shell. The 
imaginary part has been shown to lead to a small renormalization of the
precession frequencies\cite{FreqRenorm} that can only reach values over
1\% for a small range of realistic material parameters and excitation 
conditions, so that we consider only the real part of $\bar{\delta}$
in the further discussion of the Markov limit.

In the above treatment, it was postulated that the memory induced by the 
correlations is short. To see in which cases this is indeed a good 
approximation and how the timescale of the memory can be defined,
we briefly summarize the findings of Ref.~\onlinecite{Proceedings}:
The source terms $b_{l\k_1}^{\alpha\k_2}$ that enter, e.g., in the dynamics
for the carrier spin $\mbf s_{\k_1}$, involve the variables
$n_{\k_1}$, $\mbf s_{\k_1}$, $n_{\k_2}$ and $\mbf s_{\k_2}$. For the parts
that only contain variables at $\k_1$, which we will refer to as
$b_{l\k_1}^{\alpha}$, the real part of the integral on the right-hand side of 
Eq.~(\ref{eq:sumkQ}) yields:
\begin{align}
&\Re\int\limits_{-\infty}^t\! dt'\, 
\int\limits_{0}^{\omega_{BZ}}\!d\omega\,
e^{i(\omega-\omega_{\k_1})(t'-t)} 
b_{l\k_1}^{\alpha}(t')=\nn&=
\Re\int\limits_{-\infty}^0\! dt''\, 
\frac{\sin\big[(\omega_{BZ}-\omega_{\k_1})t''\big]
+\sin(\omega_{\k_1} t'')}{t''} b_{l\k_1}^{\alpha}(t+t'')
\end{align}
Since $\frac{\sin \Delta\omega t}{t}\to \pi\delta(t)$ for 
$\Delta\omega\to\infty$, this way of expressing the integral now shows that 
the memory has two timescales, one corresponding to 
$(\omega_{BZ}-\omega_{\k_1})^{-1}$, which is typically of the order of a few 
fs due to values of $\omega_{BZ}$ in the eV range, and the other one
at $\omega_{\k_1}^{-1}$. This can explain, why for a $\delta$-like 
initial electron occupation at $\k_1=0$, the spin transfer rate extracted 
from the quantum kinetic calculations in Ref.~\onlinecite{Proceedings}
is exactly $\frac 12$ of the Markovian expression for the rate. Thus,
non-Markovian effects are found to be mainly due to the spectral proximity  
of the electrons to the band edge. Therefore, if the initial carrier 
distribution has a width of a few meV, the Markovian results coincide
with the quantum kinetic calculations\cite{Proceedings}. 

For the other parts of the source terms $b_{l\k_1}^{\alpha\k_2}$ which
depend also on the electron variables at $\k_2$, a new timescale emerges
which corresponds to the inverse of the frequency difference 
$\tau_{\k_1,\k_2}$ for which the electron variables
$\mbf s_{\k_2}$ ($n_{\k_2}$) start to differ notably from $\mbf s_{\k_1}$
($n_{\k_1}$). 

In summary, it can therefore be said that the correlation time $\tau_{cor}$,
i.~e. the timescale of the memory induced by the correlations, 
depends on the details of the spectral carrier distributions. 
Thus, in order to obtain meaningful results 
by using the Markov approximation, it is of key importance
 that the dynamics of the 
source terms takes place on a much slower timescale than the build-up of 
correlations $\tau_{cor}$. 
If this is not the case, e.g., due to a fast precession of the electron 
spins with a frequency $\omega_e$, it is necessary to split this precession 
off of the correlation induced spin transfer, yielding a modified
integral kernel $e^{i(\omega_{\k_2}-\omega_{\k_1}\pm\omega_e)t'}$ 
and therefore a shift of $\pm\omega_e$ in the respective 
$\delta$-functions\cite{PESC}.
Therefore, the identification of fast and slowly changing parts of 
the source terms $b_{l\k_1}^{\alpha\k_2}$ is crucial for 
the derivation of the Markov limit of the 
quantum kinetic equations of motion~(\ref{eq:eom}).

\subsection{Markov Limit of the Quantum Kinetic Equations}

In the last section, the standard procedure of deriving a Markov limit was 
summarized starting from a simple set of equations where all the relevant
spin precessions in DMS were neglected. Now, for the more general theory
of the present article, we repeat the same steps
while accounting for all terms in Eqs.~(\ref{eq:eom}). 
As above, first of all, the homogeneous part of the differential equation
for the correlations has to be solved.
\begin{align}
&\ddt {Q_{l\k_1}^{\alpha\k_2}}^{\t{hom}}=
-i(\omega_{\k_2}-\omega_{\k_1}){Q_{l\k_1}^{\alpha\k_2}}^{\t{hom}}+\nn&
+\big(A_{\k_1}+A_{\k_2}^*\big)_{\alpha\gamma}{Q_{l\k_1}^{\gamma\k_2}}^{\t{hom}}
+\epsilon_{ijl} \omega_{Mn}^i {Q_{j\k_1}^{\alpha\k_2}}^{\t{hom}}.
\label{eq:Qhom}
\end{align}
Eq.~(\ref{eq:Qhom}) can be represented in a more abstract form, if
${Q_{l\k_1}^{\alpha\k_2}}^{\t{hom}}$ is rewritten as a single vector 
$\mbf Q^{\t{hom}}$
with respect to the set of indices $l$, $\alpha$, $\k_1$ and $\k_2$. Then, 
Eq.~(\ref{eq:Qhom}) becomes:
\begin{align}
&\ddt \mbf Q^{\t{hom}}=M \mbf Q^{\t{hom}}
\label{eq:Qvechom}
\end{align}
where the matrix $M$ is defined by the terms on the r.~h.~s. of 
Eq.~(\ref{eq:Qhom}). 
The formal solution of Eq.~(\ref{eq:Qvechom}) is the
time ordered exponential:
\begin{align}
\mbf Q^{\t{hom}}(t_0+\Delta t)=\mathcal{T}e^{\int\limits_{t_0}^{t_0+\Delta t}
dt' M(t') } \mbf Q^{\t{hom}}(t_0)
\label{eq:QvecTO}
\end{align}

However, since in the Markov limit
the solution of the homogeneous differential equation is only required
on a timescale comparable to $\tau_{cor}$ in the fs range, 
we can assume that neither the precession frequencies nor the
precession axes will change significantly on this timescale. 
This assumption makes it possible to approximate $M(t')\approx M(t_0)$ in
Eq.~(\ref{eq:QvecTO}) so that the time ordering operator $\mathcal{T}$ 
can be dropped.

The expression for the solution for $\mbf Q^{\t{hom}}$ can be further 
simplified, because the different contributions to the r.~h.~s. of 
Eq.~(\ref{eq:Qhom}) act on different degrees of freedom of 
${Q_{l\k_1}^{\alpha\k_2}}^{\t{hom}}$ and therefore commute.
As also $A_{\k_1}$ and $A_{\k_2}^*$ commute, which can be 
checked directly using the explicit expression for those matrices in 
Eq.~(\ref{eq:defA}), the homogeneous part of the equation of motion for
the correlation is solved by 
\begin{align}
&{Q_{l\k_1}^{\alpha\k_2}}^{\t{hom}}(t_0+\Delta t)=
e^{-i(\omega_{\k_2}-\omega_{\k_1})\Delta t} \times\nn&
\big(e^{A_{\k_1}\Delta t} e^{A_{\k_2}^*\Delta t}\big)_{\alpha\gamma}
\big(e^{ [\boldsymbol\omega_{Mn}]_\times\Delta t}\big)_{ll'}
{Q_{l\k_1}^{\alpha\k_2}}^{\t{hom}}(t_0)
\end{align}

The exponential $e^{ [\boldsymbol\omega_{Mn}]_\times t}$ of the 
cross product matrix $[\boldsymbol\omega_{Mn}]_\times $ is 
\begin{align}
e^{ [\boldsymbol\omega_{Mn}]_\times t}=R_{\boldsymbol\omega_{Mn}}
(\omega_{Mn}t),
\end{align}
where $R_{\mbf n}(\alpha)$ is the $3\!\!\times\!\!3$ matrix describing a
rotation around the axis $\mbf n$ with an angle $\alpha$.
Similarly, it is possible to calculate an exponential of the matrices 
$A_{\k}$: 
\begin{align}
&E_{\k_1}(t):=e^{A_{\k_1}t}=\nn&=
\cos\big(\frac{\Omega'_{\k_1}}2 t\big)\mbf 1 +
\sin\big(\frac{\Omega'_{\k_1}}2 t\big)\left(\begin{array}{cc} 
0&2i\frac{{\boldsymbol\Omega'_{\k_1}}^{\!\!T}}{\Omega'_{\k_1}}\\
\frac i2\frac{\boldsymbol\Omega'_{\k_1}}{\Omega'_{\k_1}}&
[\frac{\boldsymbol\Omega'_{\k_1}}{\Omega'_{\k_1}}]_\times
\end{array}\right),
\label{eq:Edef}
\end{align}
with the inverse $\big(E_{\k_1}(t)\big)^{-1}=E_{\k_1}(-t)$.

Now, the solution to the inhomogeneous equation can be found by a 
variation of constants yielding:
\begin{align}
&Q_{l\k_1}^{\alpha\k_2}(t_0+\Delta t)=
e^{-i(\omega_{\k_2}-\omega_{\k_1})\Delta t} 
\big(E_{\k_1}(\Delta t) E_{\k_2}^*(\Delta t)\big)_{\alpha\gamma}\times\nn&
\big(R_{\boldsymbol\omega_{Mn}}(\omega_{Mn}\Delta t)\big)_{ll'}
\bigg[Q_{l\k_1}^{\alpha\k_2}(t_0)
+\int\limits_{t_0}^{t_0+\Delta t} dt'\,
e^{i(\omega_{\k_2}-\omega_{\k_1})t'} \times\nn&
\big(E_{\k_1}(-t')E_{\k_2}^*(-t')\big)_{\gamma\kappa}
\big(R_{\boldsymbol\omega_{Mn}}(-\omega_{Mn}t')\big)_{l'l''}
b_{l''\k_1}^{\kappa\k_2}(t')\bigg]
\label{eq:Qformal}
\end{align}

Eq.~(\ref{eq:Qformal}) can be further simplified by 
decomposing the matrices $R_{\boldsymbol\omega_{Mn}}(\omega_{Mn}t)$ and
$E_{\k_1}$ as well as $E_{\k_2}^*$ in components oscillating with 
different frequencies:
\begin{subequations}
\begin{align}
&R_{\mbf n}(\omega t)=R^0_{\mbf n} +
R^{+}_{\mbf n} e^{i\omega t}+R^{-}_{\mbf n} e^{-i\omega t} \\
&E_{\k_1}(t)=E_{\k_1}^0+
E_{\k_1}^{+} e^{i\frac 12\Omega'_{\k_1}t}+
E_{\k_1}^{-} e^{-i\frac 12\Omega'_{\k_1}t}\\
&E_{\k_2}^*(t)=(E_{\k_2}^*)^0+
(E_{\k_2}^*)^{+} e^{i\frac 12\Omega'_{\k_2}t}+
(E_{\k_2}^*)^{-} e^{-i\frac 12\Omega'_{\k_2}t},
\end{align}
\label{eq:freqcomp}
\end{subequations}
where the components of $E_\k$ can directly be read off from the definition
in Eq.~(\ref{eq:Edef}) and the decomposition of $R_{\mbf n}(\alpha)$ is
\begin{subequations}
\begin{align}
&\big(R_{\mbf n}^0\big)_{ij}=\frac{n_i n_j}{|\mbf n|^2}\\
&\big(R_{\mbf n}^\pm\big)_{ij}=\frac 12\Big(\delta_{ij}
-\frac{n_i n_j}{|\mbf n|^2}\pm i\epsilon_{ijk}\frac{n_k}{|\mbf n|}\Big).
\end{align}
\label{eq:RchiDef}
\end{subequations}

For the components defined in Eq.~(\ref{eq:freqcomp}), an important 
relation is
\begin{subequations}
\begin{align}
&R^{\chi_1}_{\mbf n} R^{\chi_2}_{\mbf n}=
\delta_{\chi_1\chi_2} R^{\chi_1}_{\mbf n},\\
&E^{\chi_1}_{\k_1}E^{\chi_2}_{\k_1}=
\delta_{\chi_1\chi_2} E^{\chi_1}_{\k_1},\\
&(E_{\k_2}^*)^{\chi_1}(E_{\k_2}^*)^{\chi_2}=
\delta_{\chi_1\chi_2} (E_{\k_2}^*)^{\chi_1},
\end{align}
\label{eq:RRel}
\end{subequations}
where from now on we assume $\chi_i\in\{-1,0,1\}$ for any $\chi$-index.

As stated earlier, it is necessary to identify fast oscillating and slowly 
changing contributions to the source terms $b_{l\k_1}^{\alpha\k_2}$. 
To this end, we consider the 
dynamics of $b_{l\k_1}^{\alpha\k_2}$ in the mean-field approximation, where
\begin{subequations}
\begin{align}
\langle S^i S^j (t_0+\Delta t)\rangle\approx&
\big(R_{\boldsymbol\omega_{Mn}}(\omega_{Mn}\Delta t)\big)_{ii'}\times\nn&
\big(R_{\boldsymbol\omega_{Mn}}(\omega_{Mn}\Delta t)\big)_{jj'}
\langle S^{i'} S^{j'} (t_0)\rangle,\\
n_{\k}(t_0+\Delta t)\approx& n_{\k} (t_0),\\
s_{\k}^i(t_0+\Delta t)\approx &
\big(R_{\boldsymbol\Omega'_{\k}}(\Omega'_{\k}\Delta t)\big)_{ii'}
s^{i'}_{\k}(t_0),
\end{align}
\end{subequations}
With these approximations, the source terms can be decomposed into
\begin{align}
&b_{l\k_1}^{\alpha\k_2}(t_0+\Delta t)\approx\sum_m 
{b_{l\k_1}^{\alpha\k_2}}^{(\omega_m)}(t_0) e^{i\omega_m \Delta t}, 
\label{eq:bdecomp}
\end{align}
where $m$ counts all the possible oscillation frequencies $\omega_m$
which consist of combinations of the frequencies $\omega_{Mn}(z)$ and 
$\Omega'_{\k}$.

Now, the Markov limit of the Eqs.~(\ref{eq:eom}) can be established
by using the Markov approximation in Eq.~(\ref{eq:sumQmarkov}) with the
Sokhotski-Plemelj formula in Eq.~(\ref{eq:Plemelj}) on the expression for
the time evolution of the correlations in Eq.~(\ref{eq:Qformal}), 
simplifying the products of exponential matrices with the 
relations~(\ref{eq:RRel}) and decomposing the source terms 
according to Eq.~(\ref{eq:bdecomp}):
\begin{align}
Q_{l\k_1}^{\alpha\k_2}\approx &\pi \sum_m
\sum_{\chi_{Mn},\chi_{\k_1},\chi_{\k_2}}
\bar{\delta}\bigg(\omega_{\k_2}-\Big(\omega_{\k_1}+\chi_{Mn}\omega_{Mn}+\nn&
+\frac12\chi_{\k_1}\Omega'_{\k_1}+\frac12\chi_{\k_2}\Omega'_{\k_2}
-\omega_m\Big)\bigg)
\times\nn&
\big(E_{\k_1}^{\chi_{\k_1}} (E_{\k_2}^*)^{\chi_{\k_2}}\big)_{\alpha\gamma}
\big(R_{\boldsymbol\omega_{Mn}}^{\chi_{Mn}}\big)_{ll'}
{b_{l'\k_1}^{\gamma\k_2}}^{(\omega_m)}(t')
\end{align}
or more explicitly:

\begin{widetext}
\begin{align}
&Q_{l\k_1}^{\alpha\k_2}(z)\approx\pi\frac i\hbar \Jsd |\psi(z)|^2 d
\sum_{\chi_{\k_1},\chi'_{\k_1},\chi_{\k_2},\chi'_{\k_2},\chi_{Mn}} \bar{\delta}\bigg(
\omega_{\k_2}-\Big(\omega_{\k_1}+\big(\frac12\chi_{\k_1}-\chi'_{\k_1}\big)\Omega'_{\k_1}+
\big(\frac 12 \chi_{\k_2}-\chi'_{\k_2}\big)\Omega'_{\k_2}
-\chi_{Mn}\omega_{Mn}(z) \Big)\bigg)\bigg\{\nn&
\big(E_{\k_1}^{\chi_{\k_1}}(E_{\k_2}^*)^{\chi_{\k_2}}\big)_{\alpha 0}\Bigg[
\frac{\langle S^{l}S^{j'}(z)+S^{j'}S^{l}(z)\rangle}2
\big(R_{\boldsymbol\omega_{Mn}(z)}^{\chi_{Mn}}\big)_{jj'}\Big[
\delta_{\chi'_{\k_1},0}
\big(R_{\boldsymbol\Omega'_{\k_2}}^{\chi'_{\k_2}}\big)_{jk'} s_{\k_2}^{k'}
-\delta_{\chi'_{\k_2},0}
\big(R_{\boldsymbol\Omega'_{\k_1}}^{\chi'_{\k_1}}\big)_{jk'}s_{\k_1}^{k'}\Big] 
+\nn&+
\frac i2 \epsilon_{j'li''}\langle S^{i''}(z)\rangle
\big(R_{\boldsymbol\omega_{Mn}(z)}^{\chi_{Mn}}\big)_{jj'}
\Big[
\delta_{\chi'_{\k_1},0} (1-n_{\k_1})
\big(R_{\boldsymbol\Omega'_{\k_2}}^{\chi'_{\k_2}}\big)_{jk'}s_{\k_2}^{k'}
+\delta_{\chi'_{\k_2},0} (1-n_{\k_2})
\big(R_{\boldsymbol\Omega'_{\k_1}}^{\chi'_{\k_1}}\big)_{jk'}s_{\k_1}^{k'}
-2i\epsilon_{jki}
\big(R_{\boldsymbol\Omega'_{\k_1}}^{\chi'_{\k_1}}\big)_{kk'}
\big(R_{\boldsymbol\Omega'_{\k_2}}^{\chi'_{\k_2}}\big)_{ii'}
s_{\k_1}^{k'}s_{\k_2}^{i'}\Big]
\Bigg]
+\nn&+
\big(E_{\k_1}^{\chi_{\k_1}}(E_{\k_2}^*)^{\chi_{\k_2}}\big)_{\alpha k}\Bigg[
\big(R_{\boldsymbol\omega_{Mn}(z)}^{\chi_{Mn}}\big)_{kj'}
\delta_{\chi'_{\k_1},0}\delta_{\chi'_{\k_2},0}
\bigg[\frac{\langle S^{l}S^{j'}(z)+S^{j'}S^{l}(z)\rangle}2
\frac{n_{\k_2}-n_{\k_1}}4
+\frac i2\epsilon_{j'li''}\langle S^{i''}(z)\rangle
\Big(\frac{n_{\k_2}+n_{\k_1}-n_{\k_1}n_{\k_2}}4\Big)\bigg]+\nn&+
\frac i2\epsilon_{j'li''}\langle S^{i''}(z)\rangle
(\delta_{jk}\delta_{k'k''}-\delta_{jk'}\delta_{kk''}-\delta_{jk''}\delta_{kk'})
\big(R_{\boldsymbol\omega_{Mn}(z)}^{\chi_{Mn}}\big)_{jj'}
\big(R_{\boldsymbol\Omega'_{\k_1}}^{\chi'_{\k_1}}\big)_{k'i}
\big(R_{\boldsymbol\Omega'_{\k_2}}^{\chi'_{\k_2}}\big)_{k''i'}
s_{\k_1}^i s_{\k_2}^{i'}+\nn&+
\frac i2\epsilon_{jki}\frac{\langle S^{l}S^{j'}(z)+S^{j'}S^{l}(z)\rangle}2
\big(R_{\boldsymbol\omega_{Mn}(z)}^{\chi_{Mn}}\big)_{jj'}
\Big[
\big(R_{\boldsymbol\Omega'_{\k_2}}^{\chi'_{\k_2}}\big)_{il'}
\delta_{\chi'_{\k_1},0} s_{\k_2}^{l'}
+\big(R_{\boldsymbol\Omega'_{\k_1}}^{\chi'_{\k_1}}\big)_{il'}
\delta_{\chi'_{\k_2},0} s_{\k_1}^{l'}
\Big]
+\nn&
-\frac 14\epsilon_{jki}\epsilon_{j'li''}\langle S^{i''}(z)\rangle
\big(R_{\boldsymbol\omega_{Mn}(z)}^{\chi_{Mn}}\big)_{jj'}
\Big[
\big(R_{\boldsymbol\Omega'_{\k_2}}^{\chi'_{\k_2}}\big)_{il'}
\delta_{\chi'_{\k_1},0}(1- n_{\k_1}) s_{\k_2}^{l'}
-
\big(R_{\boldsymbol\Omega'_{\k_1}}^{\chi'_{\k_1}}\big)_{il'}
\delta_{\chi'_{\k_2},0}(1- n_{\k_2}) s_{\k_1}^{l'}\Big]
\Bigg]\bigg\},
\label{eq:QkdepMarkov}
\end{align} 
\end{widetext}
Finally, inserting the expression for $Q_{l\k_1}^{\kappa\k_2}$ of
Eq.~(\ref{eq:QkdepMarkov}) into the quantum kinetic 
equations of motion (\ref{eq:eom}a)-(\ref{eq:eom}c) 
for the carrier and impurity variables, yields the desired
set of ordinary differential equations for $n_{\k}$, $\mbf s_{\k}$ 
and $\langle\mbf S\rangle$
where the correlations are eliminated, but their effects are still 
accounted for.

\subsection{Numerical Implementation of the Markovian Equations of motion}
The numerical advantage of the Markov limit over the original 
quantum kinetic equations is mainly that, because of the 
$\delta$-function in Eq.~(\ref{eq:QkdepMarkov}), only those electronic
states with wave vectors $\k_2$ contribute to the time evolution of 
electron variables with wave vector $\k_1$ that are allowed by
energy conservation.
Here, the total energy consists of the kinetic energy as 
well as Zeeman-like spin-dependent energies due to the impurity magnetization,
the external magnetic field and the $\k$-dependent effective magnetic field
due to the Rashba- or Dresselhaus-terms as well as the impurity Zeeman energy.

The complicated interplay of the different contributions to the total energy
makes it hard to find the
roots of the argument of the 
$\delta$-function in Eq.~(\ref{eq:QkdepMarkov}), which is
necessary in order to identify the wave vectors $\k_2$ of the 
electronic states which are relevant for the calculation 
of the time evolution of electronic states with wave vector $\k_1$.
In particular, the $\k$-dependence of the energies, the
dimensionality of the $\k$-vector and the fact that the number of roots
is in general not known turn out to be major obstacles for the direct numerical
solution of Eq.~(\ref{eq:QkdepMarkov}).

Here, we solve this problem by rediscretizing the electron variables.
The roots of the argument of the $\delta$-function in 
Eq.~(\ref{eq:QkdepMarkov}) are given by:
\begin{subequations}
\begin{align}
\bar{\omega}_{\k_2}(\xi_2)=\bar{\omega}_{\k_1}(-\xi_1)-\chi_{Mn}\omega_{Mn}(z)
\end{align}
with
\begin{align}
&\bar{\omega}_{\k}(\xi):=\omega_{\k}-\xi \Omega'_{\k},\\
&\xi\in\{-\frac 32,-\frac 12,\frac 12,\frac 32\}
\end{align}
\end{subequations}
After the space of $\bar{\omega}$ is discretized into small intervals,
we create a list of discretization points in $\k$-space which contribute
to the corresponding interval with respect to $\bar{\omega}$.
Since the construction of this list has the complexity $\mathcal{O}(N_k^2)$,
where $N_k$ is the number of discretization points of a linear dimension in
the two-dimensional $\k$-space, and the correlations 
$\sum_{\k_2}Q_{l\k_1}^{\alpha\k_2}$ which enter in the equation for 
a single electron variable with wave-vector $\k_1$ become of the order 
of $\mathcal{O}(N_k^0)$ due to the $\delta$-function, the problem 
of solving the Markovian equations in the full $\k$-space is 
$\mathcal{O}(N_k^2)$.
This provides a significant advantage over the full quantum kinetic 
theory which has the complexity $\mathcal{O}(N_k^4)$ for a quantum well.

\subsection{Case $\NMn\gg N_e$ without spin-orbit fields}
The Markov limit (\ref{eq:QkdepMarkov}) of the equations of 
motion (\ref{eq:eom}) yields quite lengthy expressions. However, these
can be simplified dramatically in a case which is very common for
experimentally studied DMS samples: If the number of the magnetic
impurites $\NMn$ exceeds largely the number of quasi-free carriers $N_e$,
such as in the case of optically excited intrinsic DMS, the impurity spin
$\langle \mbf S\rangle$ only changes marginally due to the influence from
the quasi-free carriers. 
One can therefore assume that the impurity spin will approximately 
be defined by its thermal equilibrium value in the external magnetic field.
In particular in the paramagnetic regime, the impurity spin will
be parallel $(\sigma^B_S=+1)$ or anti-parallel $(\sigma^B_S=-1)$
to the magnetic field
\begin{subequations}
\begin{align}
&\langle \mbf S\rangle=\sigma^B_S|\langle \mbf S\rangle|\frac{\mbf B}{|\mbf B|}.
\end{align}
Since usually the Zeeman contribution to the energy of the magnetic ions
is much stronger than the mean-field $s$-$d$ term due to the carrier 
spins\cite{ColMod2}, we assume that
\begin{align}
&\boldsymbol\omega_{Mn}=\sigma^B_{Mn} \omega_{Mn}  \frac{\mbf B}{|\mbf B|}
\end{align}
and that $\boldsymbol\omega_{Mn}$ is independent of $z$.
If only electrons with small wave vectors are excited, 
no electric field is applied, and the sample has a rather 
high impurity concentration,
the $s$-$d$ interaction usually dominates over spin-orbit coupling effects, 
so that one can neglect the latter\cite{Ungar15}. 
Here, we shall first concentrate on this case and defer the
discussion of the interplay between s-d interactions and
spin-orbit coupling to section \ref{sdSO}.
Since the external magnetic field 
as well as the effective $s$-$d$ field due the impurity spins are 
parallel, we find also
\begin{align}
&\boldsymbol\Omega_\k=
\boldsymbol\omega_{e}=\sigma^B_{e} \omega_{e}  \frac{\mbf B}{|\mbf B|}.
\end{align}
\label{eq:cond_par}
\end{subequations}

Because of the $\k$-independence of the effective magnetic field for the 
carriers, the matrix $E_{\k_1}E_{\k_2}^*$ can be simplified to
\begin{align}
&E_{\k_1}E_{\k_2}^*= \left(\begin{array}{cc}1&0\\0&
R_{\boldsymbol\omega_e} \end{array}\right). 
\end{align}
Additionally, comparing Eq.~(\ref{eq:Ssec}) with Eq.~(\ref{eq:RchiDef})
yields
\begin{align}
\langle S^{i'}S^{j'}\rangle=&
\langle{S^\|}^2\rangle \big(R^0_{\langle \mbf S\rangle}\big)_{i'j'}+
\langle{S^\perp}^2\rangle 
\big(R^+_{\langle \mbf S\rangle}+R^-_{\langle \mbf S\rangle}\big)_{i'j'}
+\nn&
+|\langle \mbf S\rangle| \frac 12
\big(R^+_{\langle \mbf S\rangle}-R^-_{\langle \mbf S\rangle}\big)_{i'j'}.
\end{align}

Now, the products of matrices in Eq.~(\ref{eq:QkdepMarkov}) 
can be evaluated using 
\begin{align}
&R_{\langle\mbf S\rangle}^{\sigma^B_S \chi}=
R_{\boldsymbol\omega_{Mn}}^{\sigma^B_{Mn} \chi}=
R_{\boldsymbol\omega_e}^{\sigma^B_e\chi} =
R_{\mbf B}^\chi,
\end{align}
and the relation (\ref{eq:RRel}a).
After a lengthy but straightforward calculation, we arrive at
the Markov limit of the equations of motion for the occupations 
of the spin-up and -down subbands with respect to the direction of
the external magnetic field
$n^{\up/\down}_{\k}:=\frac{n_{\k}}2\pm \frac{\mbf B}{|\mbf B|}\cdot\mbf s_{\k}$ 
and the perpendicular spin component
$\mbf s^\perp_{\k}:=\mbf s_{\k}- \frac{\mbf B}{|\mbf B|}\big(
\frac{\mbf B}{|\mbf B|}\cdot\mbf s_{\k}\big)$:
\begin{widetext}
\begin{subequations}
\begin{align}
\ddt n^{\up/\down}_{\k_1}\big|_{cor}\approx& \int\limits_{-\frac d2}^{\frac d2}dz\,
\pi\frac{\Jsd^2|\psi(z)|^4\NMn d}{\hbar^2 V^2}\sum_{\k_2}
\bigg\{ \delta(\omega_{\k_2}-\omega_{\k_1})\frac{\langle{S^\|}^2\rangle}2
(n_{\k_2}^{\up/\down}-n_{\k_1}^{\up/\down})
+\delta\Big(\omega_{\k_2}-\big(\omega_{\k_1}\pm(\sigma^B_e\omega_e
-\sigma^B_e\omega_{Mn})\big)\Big)\times\nn&\Big[
\Big(\langle{S^\perp}^2\rangle \pm\sigma^B_S\frac{|\langle\mbf S\rangle|}2\Big)
(1-n_{\k_1}^{\up/\down}) n_{\k_2}^{\down/\up}-
\Big(\langle{S^\perp}^2\rangle \mp\sigma^B_S\frac{|\langle\mbf S\rangle|}2\Big)
(1-n_{\k_2}^{\down/\up}) n_{\k_1}^{\up/\down}\Big]\bigg\},\\
\ddt \mbf s^\perp_{\k_1}\big|_{cor}\approx& \int\limits_{-\frac d2}^{\frac d2}dz\,
\pi\frac{\Jsd^2|\psi(z)|^4\NMn d}{\hbar^2 V^2}\sum_{\k_2}
\bigg\{-\delta(\omega_{\k_2}-\omega_{\k_1})\frac{\langle{S^\|}^2\rangle}2
(\mbf s_{\k_2}^{\perp}+\mbf s_{\k_1}^{\perp})+\nn&
-{\delta}\Big(\omega_{\k_2}-\big(\omega_{\k_1}+(\sigma^B_e\omega_e
-\sigma^B_e\omega_{Mn})\big)\Big) \Big[\frac 12
\Big(\langle{S^\perp}^2\rangle -\sigma^B_S\frac{|\langle\mbf S\rangle|}2\Big) 
+n_{\k_2}^{\up}\sigma^B_S\frac{|\langle\mbf S\rangle|}2\Big]
\mbf s^\perp_{\k_1}+\nn&
-{\delta}\Big(\omega_{\k_2}-\big(\omega_{\k_1}-(\sigma^B_e\omega_e
-\sigma^B_e\omega_{Mn})\big)\Big) \Big[ \frac 12
\Big(\langle{S^\perp}^2\rangle +\sigma^B_S\frac{|\langle\mbf S\rangle|}2\Big) 
-n_{\k_2}^{\down}\sigma^B_S\frac{|\langle\mbf S\rangle|}2\Big]
\mbf s^\perp_{\k_1}+\nn&
-\frac 1\pi\frac 1{\omega_{\k_2}-\big(\omega_{\k_1}+(\sigma^B_e\omega_e
-\sigma^B_e\omega_{Mn})\big)}\Big[\frac 12
\Big(\langle{S^\perp}^2\rangle -\sigma^B_S\frac{|\langle\mbf S\rangle|}2\Big) 
+n_{\k_2}^{\up}\sigma^B_S\frac{|\langle\mbf S\rangle|}2\Big]
\frac{\mbf B}{|\mbf B|}\times \mbf s^\perp_{\k_1}+\nn&
+\frac 1\pi\frac 1{\omega_{\k_2}-\big(\omega_{\k_1}-(\sigma^B_e\omega_e
-\sigma^B_e\omega_{Mn})\big)}\Big[ \frac 12
\Big(\langle{S^\perp}^2\rangle +\sigma^B_S\frac{|\langle\mbf S\rangle|}2\Big) 
-n_{\k_2}^{\down}\sigma^B_S\frac{|\langle\mbf S\rangle|}2\Big]
\frac{\mbf B}{|\mbf B|}\times \mbf s^\perp_{\k_1}
\bigg\},
\end{align}
\label{eq:PESC}
\end{subequations}
\end{widetext}
where $\ddt n^{\up/\down}_{\k_1}\big|_{cor}$ and 
$\ddt \mbf s^\perp_{\k_1}\big|_{cor}$ describe the contributions to the 
time derivative of the respective quantities beyond the mean-field 
dynamics. In the case studied here, the total time evolution is given by:
\begin{subequations}
\begin{align}
&\ddt n^{\up/\down}_{\k_1}=\ddt n^{\up/\down}_{\k_1}\big|_{cor},\\
&\ddt \mbf s^\perp_{\k_1}=\boldsymbol{\omega}_e\times\mbf s^\perp_{\k_1}+
\ddt \mbf s^\perp_{\k_1}\big|_{cor},\\
&\ddt \langle\mbf S\rangle=\boldsymbol{\omega}_{Mn}\times\langle\mbf S\rangle
+\ddt \langle\mbf S\rangle\big|_{cor},
\end{align}
\label{eq:PESCMF}
\end{subequations}
where $\ddt \langle\mbf S\rangle\big|_{cor}$ can be obtained by replacing
$\NMn\int\limits_{-\frac d2}^{\frac d2}dz$ by 
$-d\sum\limits_{\k_1}$ in the r.~h.~s. of Eq.~(\ref{eq:PESC}b).
This follows directly from the fact that the $s$-$d$ interaction conserves
the total spin. 

Note that Eqs.~(\ref{eq:PESC}) generalize Eqs. (6) of Ref.~\onlinecite{PESC} 
by incorporating a $\k$-dependent precession frequency for the electrons, 
an external magnetic field and the z-dependence of the coupling due to the 
form of the envelope function of the quantum well.

Eq.~(\ref{eq:PESC}a) can be interpreted like equations resulting from
Fermi's golden rule: A spin-up electron is scattered either to another 
spin-up state with the same value of the kinetic energy $\hbar\omega_\k$
(term proportional to $\delta(\omega_{\k_2}-\omega_{\k_1})$) or
to a spin-down state with kinetic energy $\hbar\omega_{\k_2}=
\hbar\omega_{\k_1}+\hbar(\sigma^B_e\omega_e-\sigma^B_e\omega_{Mn})$
and vice-versa. To understand the latter term it is important to 
keep in mind that the total mean-field energy of a spin-up electron is 
$\hbar(\omega_\k+\frac 12\sigma^B_e\omega_e)$ while for a spin-down electron
one finds $\hbar(\omega_\k-\frac 12\sigma^B_e\omega_e)$. Also, since the
$s$-$d$ interaction conserves the sum of the electron and impurity spins,
a flip of an electron spin in one direction is always accompanied by a
flip of an impurity spin in the opposite direction. Thus, in order to fulfill 
the conservation of the total mean-field energy, the energy
$\hbar(\sigma^B_e\omega_e-\sigma^B_e\omega_{Mn})$ that is freed by an
impurity mediated flip of an electron from the spin-up to the spin-down state
has to be compensated by a difference of the 
kinetic energies of the electronic states $\omega_{\k_2}-\omega_{\k_1}$. 

Although Eq.~(\ref{eq:PESC}a) for the spin-up and spin-down occupations
can also be derived by Fermi's golden rule, 
the energy shifts in the $\delta$-functions are often not correctly accounted
for in the literature\cite{Wu09,Semenov}. 
The consequences are discussed in section \ref{Res_Bdep}.
Here, the spin-flip terms of Eq.~(\ref{eq:PESC}a) also correctly 
account for Pauli-blocking effects by the terms proportional to
$(1-n_{\k}^{\up/\down})$ which are usually put in by hand in a golden rule
derivation.
Furthermore, a golden rule treatment only allows to derive transition rates 
between energy eigenstates and does not provide equations governing the 
dynamics of the coherences between those eigenstates, i.~e. the components of 
the electron and impurity spins perpendicular to the direction of the external
magnetic field, which is given in our derivation by Eq.~(\ref{eq:PESC}b).
As in the equations for the spin-up and spin-down occupations, 
we find that the equations for the perpendicular spin components 
connect states whose difference in kinetic 
energies $\hbar(\omega_{\k_2}-\omega_{\k_1})$ is either zero 
or $\pm  \hbar(\sigma^B_e\omega_e-\sigma^B_{Mn}\omega_{Mn})$.
Note that in contrast to the equations for $n^{\up/\down}_{\k_1}$, here,
we find terms proportional to the imaginary part of $\bar{\delta}$. While
the real part leads to a rate-like damping of the perpendicular electron spin,
the imaginary part yields an additional contribution to the precession 
frequency. Such frequency renormalizations have been extensively discussed in 
Ref.~\onlinecite{FreqRenorm}. 

From Eqs.~(\ref{eq:PESC}) one can also find decay rates for spin-up
($\big(\tau^\up_{\k_0}\big)^{-1}$) and spin-down 
($\big(\tau^\down_{\k_0}\big)^{-1}$) electron states as well as
the spin components parallel ($\big(\tau^\|_{\k_0}\big)^{-1}$)
perpendicular ($\big(\tau^\perp_{\k_0}\big)^{-1}$)
to the external magnetic field,
if it is assumed that only very few quasi-free carriers are excited, so
that one can regard only single electrons by setting 
$n^{\up/\down}_{\k_2}=\delta_{\omega_{\k_1},\omega_{\k_2}}n^{\up/\down}_{\k_1}$
and $\mbf s^\perp_{\k_2}=\delta_{\omega_{\k_1},\omega_{\k_2}}\mbf s^\perp_{\k_1}$:

\begin{subequations}
\begin{align}
&\big(\tau^\up_{\k_0}\big)^{-1}=\Gamma^- \Theta
\big(\omega_{\k_0}+(\sigma^B_e\omega_e-\sigma^B_{Mn}\omega_{Mn})\big),\\
&\big(\tau^\down_{\k_0}\big)^{-1}=\Gamma^+ \Theta
\big(\omega_{\k_0}-(\sigma^B_e\omega_e-\sigma^B_{Mn}\omega_{Mn})\big),\\
&\big(\tau^\|_{\k_0}\big)=\big(\tau^\up_{\k_0}\big)^{-1}+
\big(\tau^\down_{\k_0}\big)^{-1}\\
&\big(\tau^\perp_{\k_0}\big)^{-1}=\Gamma^0+\frac 12
\Big[\big(\tau^\up_{\k_0}\big)^{-1}+\big(\tau^\down_{\k_0}\big)^{-1}\Big],
\end{align}
with
\begin{align}
&\Gamma^0=I \pi D^{2D}\frac{\Jsd^2\NMn}{\hbar^2V^2}\langle{S^\|}^2\rangle,\\
&\Gamma^\pm=I \pi D^{2D}\frac{\Jsd^2\NMn}{\hbar^2V^2}
\Big(\langle{S^\perp}^2\rangle\pm\sigma^B_S \frac{|\langle \mbf S\rangle|}2\Big)
,\\
&I=d\int\limits_{-\frac d2}^{\frac d2}dz\, |\psi(z)|^4,
\end{align}
\label{eq:rate_expl}
\end{subequations}
where $\Theta(x)$ is the Heaviside step function.

Thus, the main effect of the frequency shifts due to the precession of
the correlations is the opening and closing of decay channels due to
the corresponding step functions which originate from the step of the 
two-dimensional density of states at $\omega_{\k=0}$. 

\section{Results}
\subsection{Magnetic Field Dependence of the Spin Transfer Rates}
\label{Res_Bdep}
Now, we compare the theory derived in the present paper with 
the different treatments of the $s$-$d$ interaction presented by 
other groups. To this end, we focus on the case without spin-orbit interactions
and $\NMn\gg N_e$, so that the correlation induced changes in the 
carrier variables can be described by Eqs.~(\ref{eq:PESC}).
Often in the literature rates for the carrier-impurity spin transfer 
dynamics are obtained from Fermi's golden rule\cite{KossutRate3D,
TsitsishviliExcDMS,TsitsishviliSpinConsHoles,Koenig}. In 
two-dimensional systems one finds in absence of magnetic fields:
\begin{subequations}
\begin{align}
\ddt s^i_{\omega_1}=&-I\pi\frac{\Jsd^2}{\hbar^2}\frac{\NMn}{V^2}
\frac{2}3(S(S+1)) \times\nn&\int d\omega\,D^{2D}(\omega)
\delta(\omega-\omega_1) s^i_\omega=
-\tau_{FGR}s^i_{\omega_1},\\
\tau_{FGR}=&I\pi\frac{\Jsd^2}{\hbar^2}\frac{\NMn}{V^2}\frac 23(S(S+1))
\frac{Am^*}{2\pi\hbar}
\end{align}
\end{subequations}
where we assume isotropy so that the carrier spin variables are independent of 
the angle of the wave vector and can equivalently be described 
by $s^i_{|\k|}$ or $s^i_{\omega}$, with the kinetic energy
$\hbar\omega=\frac{\hbar^2|\k|^2}{2m}$. 
$\tau_{FGR}$ is Fermi's golden rule spin-transfer rate at $B=0$.
In contrast, if an external magnetic field is applied, the conduction band
is energetically split by $\sigma^B_e \omega_e$. 
This leads to the appearance of an additional energy offset
in the $\delta$-function. In our treatment, we also find an 
energy offset corresponding to the impurity Zeeman splitting
$\sigma^B_{Mn}\omega_{Mn}$ 
which is necessary for the simultaneous conservation of the total 
carrier and impurity energy as well as the total spin. 
Furthermore, Fermi's golden rule is only able to predict transitions
between energy eigenstates, whereas it makes no statement about the 
transfer of the carrier spin components perpendicular to the quantization 
axis. The distinction between parallel and perpendicular components does not 
arise for $B=0$, since in this case all directions are equivalent.
Additionally, the factor $S(S+1)$ has to be modified 
in the presence of a magnetic field that causes a non-zero 
paramagnetic impurity magnetization.

In particular, the energetic offset caused by the impurity Zeeman splitting 
is often overlooked in studies based on the golden rule 
approach\cite{TsitsishviliExcDMS,Wu09}. In Ref.~\onlinecite{Wu09}, which
is based on the kinetic spin Bloch equations (KSBEs), even
the band splitting $\sigma^B_e \omega_e$ is disregarded, but the magnetic
field dependence of the second moments of the impurity spin, 
which enters in the rates, was kept.
There are also studies\cite{TsitsishviliSpinConsHoles,Koenig,Morandi10_2} 
that explicitly include 
the band splitting as well as the impurity Zeeman 
terms, but since there the rates are derived by Fermi's golden rule, no
expression for the perpendicular spin transfer component was given.

In this context, one particularly notable theoretical derivation of 
magnetic field dependent carrier-impurity spin transfer rates was 
given by {\it Semenov} in Ref.~\onlinecite{Semenov}, which
is based on a projection operator method.
There, the electron spins are treated as a subsystem which interacts
with a bath of impurity ions. 
In Ref.~\onlinecite{Semenov}, it was assumed that the electron 
density matrix can be factorized into one part accounting for the spin 
degree of freedom and the $\k$-dependent part, which is described by
a Fermi distribution. Tracing out the $\k$-dependent part of the carrier
density matrix as well as the impurity system, rates were obtained 
for the spin degree of freedom of the carriers.
In contrast to the theory of the present article, where only 
energetic shifts associated with the spin flip-flop processes of the form
$|\sigma^B_e\omega_e-\sigma^B_{Mn}\omega_{Mn}|$ appear, the projection
operator method of Ref.~\onlinecite{Semenov} also finds terms
proportional to $|\sigma^B_e\omega_e+\sigma^B_{Mn}\omega_{Mn}|$.
As mentioned earlier, such energy shifts are in conflict with the
conservation of the total carrier and impurity energy. 
We trace the appearance of the energy non-conserving 
terms in Ref.~\onlinecite{Semenov} back to the fact that, there, 
only the positive frequency component of the electron spin precession 
was regarded, whereas the negative frequency component explicitly shows up
in the theory of the present article and leads to a cancellation of 
terms in the expression for the correlations which oscillate
with $\pm(\sigma^B_e\omega_e+\sigma^B_{Mn}\omega_{Mn})$.

\begin{figure}[t!]
\includegraphics{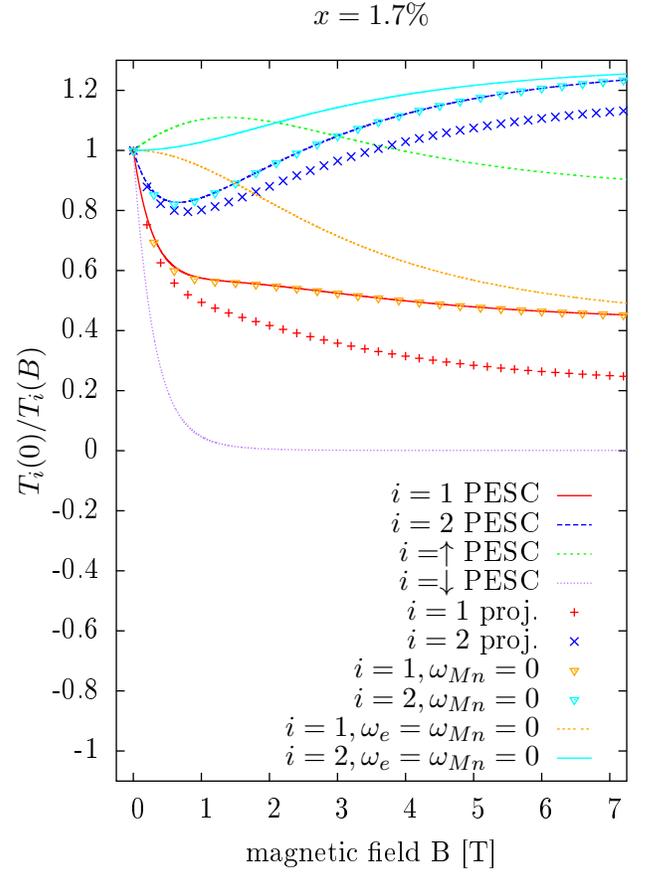}
\caption{Magnetic field dependence of the parallel ($i=1$) and
perpendicular ($i=2$) spin transfer rates 
normalized with respect to $B=0$
in a 8 nm wide Cd$_{0.983}$Mn$_{0.017}$Te 
quantum well at temperature $T=4$ K.
Red and blue lines (PESC) represent rates 
according to the theory of the present article [Eqs.~(\ref{eq:rate_comp_sem})]
and red and blue crosses show the rates calculated by the projection 
operator method (proj.) of Ref.~\onlinecite{Semenov}. 
Furthermore, cyan and orange triangles and lines show the results 
of Eqs.~(\ref{eq:rate_comp_sem}), when the energetic shifts due 
to the Zeeman impurity splittings in spin flip-flop processes 
($\omega_{Mn}=0$) or additionally the spin-splittings 
($\omega_e=\omega_{Mn}=0$) are neglected.
$T_\up$ and $T_\down$ are the relaxation rates of spin-up and 
spin-down occupations, respectively.
}
\label{fig:sem}
\end{figure}

Having discussed the different expressions for the magnetic field dependence
of the carrier-impurity spin transfer rates that can be found in the 
literature, we compare them at the example of the situation discussed in 
Ref.~\onlinecite{Semenov}. There, it was assumed that the spectral 
electron distribution is 
\begin{align}
n^\up(\omega)=n^\down(\omega)\propto e^{-\frac{\omega}T}
\end{align}
for some carrier temperature $T$, irrespective of the spin-split subband.
With this assumption, the decay rate of the total parallel ($T_1^{-1}$) and
perpendicular ($T_2^{-1}$) carrier spin with respect to the magnetic field
direction can be obtained from Eqs.~(\ref{eq:rate_expl}) of the present 
theory:
\begin{subequations}
\begin{align}
&T_1^{-1}\propto \int\limits_0^{\infty} d\omega\, e^{-\frac{\omega}T} 
\big(\tau^\|(\omega)\big)^{-1}\propto
T_\up^{-1}+T_\down^{-1} \\
&T_2^{-1}\propto \int\limits_0^{\infty} d\omega\, e^{-\frac{\omega}T} 
\big(\tau^\perp(\omega)\big)^{-1}\propto
\Gamma^0+ \frac 12\big(T_\up^{-1}+T_\down^{-1}\big) \\
&T_\up^{-1}\propto \int\limits_0^{\infty} d\omega\, e^{-\frac{\omega}T} 
\big(\tau^\up(\omega)\big)^{-1}\propto
\Gamma^-  \min( 1 , e^{\frac{\sigma^B_e\omega_e-\sigma^B_{Mn}\omega_{Mn}}T})
\\
&T_\down^{-1}\propto \int\limits_0^{\infty} d\omega\, e^{-\frac{\omega}T} 
\big(\tau^\down(\omega)\big)^{-1}\propto
\Gamma^+ \min( 1 , e^{-\frac{\sigma^B_e\omega_e-\sigma^B_{Mn}\omega_{Mn}}T})
\end{align}
\label{eq:rate_comp_sem}
\end{subequations}
where also the values for the decay rate of the
spin-up ($T_\up^{-1}$) and spin-down occupations ($T_\down^{-1}$) 
are given explicitly.
For $B=0$, the rates $T_1^{-1}=T_2^{-1}=2T^{-1}_{\up}=2T^{-1}_{\down}$ 
coincide with the rate calculated by Fermi's golden rule $\tau_{FGR}$, 
which defines the normalization of the rates in Eq.~(\ref{eq:rate_comp_sem}). 

\begin{figure}[t!]
\includegraphics{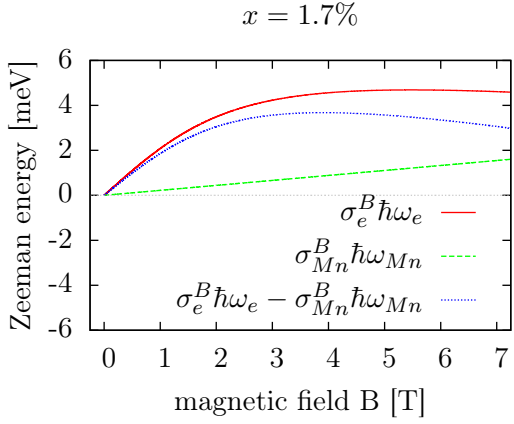}
\caption{Magnetic field dependence of the electron (red solid line) 
and impurity Zeeman energy (green dashed line) 
as well as their difference (blue dotted line)
for a DMS quantum well (same parameters as for
Fig.~\ref{fig:sem}).
}
\label{fig:sem_w}
\end{figure}

Fig.~\ref{fig:sem} shows the magnetic field dependence of the parallel and
perpendicular spin transfer rates according to Eqs.~(\ref{eq:rate_comp_sem})
with the parameters of Ref.~\onlinecite{Semenov}, where a $d=8$ nm wide 
Cd$_{0.983}$Mn$_{0.017}$Te quantum well was considered at $T=4$ K.
The value for the coupling constant is $\Jsd=15$ meVnm$^3$ and the 
electron and Mn g-factors are $g_e=-1.77$ and $g_{Mn}=2.0$ respectively.
The present theory predicts that the parallel spin transfer rate $T_1^{-1}$
first decays fast from $B=0$ to $B\approx 1$ T, then levels off. 
The perpendicular spin transfer rate $T_2^{-1}$ first decays with increasing
magnetic field, reaches a minimum at $B\approx 1$ T and finally increases 
again. 
This behaviour of $T_1^{-1}$ and $T_2^{-1}$ can be explained by considering
the rates $T^{-1}_{\up}$ and $T^{-1}_{\down}$ separately, together with
the values of the energy shifts $\sigma^B_e\omega_e-\sigma^B_{Mn}\omega_{Mn}$ 
presented in Fig.~\ref{fig:sem_w}. The mean-field impurity energy
$\hbar\sigma^B_e\omega_{Mn}$ is mainly dominated by its Zeeman energy and 
therefore increases linearly with $B$. In contrast, the mean-field carrier 
energy $\hbar\sigma^B_e\omega_e$
is strongly modified by a contribution proportional to a $S=\frac 52$ 
Brillouin-function due to the impurity magnetization, 
which starts linearly in $B$ but begins to
saturate at $B\approx 2$ T. For high magnetic fields ($B>6$ T), 
$\hbar\sigma^B_e\omega_e$ decreases again, when the impurity magnetization is 
fully saturated and the negative electron g-factor becomes important.
Although $\sigma^B_e\omega_e-\sigma^B_{Mn}\omega_{Mn}$ eventually becomes
negative for very high magnetic fields (not shown in Fig.~\ref{fig:sem_w}),
for typical experimentally accessible fields, it is mostly positive and 
increases linearly up to $B\approx 2$ T, just like $\sigma^B_e\omega_e$.

It follows from Eq.~(\ref{eq:rate_comp_sem}c) that 
$T^{-1}_\down$ decreases approximately 
exponentially with $B$ in the regime where 
$\sigma^B_e\omega_e-\sigma^B_{Mn}\omega_{Mn}$ increases linearly. Therefore,
we find that the spin-splitting introduced by the external magnetic field
closes the transfer channel $T^{-1}_\down$. In the case studied here, 
the magnetic field dependence of the rate $T^{-1}_\up$ comes exclusively from
the prefactor, since due to the positive value 
of $\sigma^B_e\omega_e-\sigma^B_{Mn}\omega_{Mn}$ the corresponding transfer
channel is maximally open. Noting that
\begin{subequations}
\begin{align}
&\Gamma^0(B\to\infty)\to\frac{15}{14}\tau_{FGR},\\
&\Gamma^+(B\to\infty)\to0,\\
&\Gamma^-(B\to\infty)\to \frac{3}{7}\tau_{FGR},
\end{align}
\end{subequations}
we find that $T^{-1}_1$ approaches $\frac{5}{8}\tau_{FGR}$ and 
$T^{-1}_2\to\frac{9}{7}\tau_{FGR}\approx 1.29\tau_{FGR}$ 
for large values of $B$.

The magnetic field dependence of rates predicted from the 
projection operator method of Ref.~\onlinecite{Semenov} is
qualitatively similar to that of the present theory, as can be seen in 
Fig.~\ref{fig:sem}. 
However, they suggest quantitatively smaller values for the rates,
with deviations of the order of $\sim 0.2 \tau_{FGR}$.
In the case studied here, the offset due to the impurity Zeeman splitting
$\sigma^B_{Mn}\omega_{Mn}$ plays a less significant role, so that 
the rates calculated neglecting these terms 
(triangles in Fig.~\ref{fig:sem}) coincide with the caculation which
conserves the total energy. However, neglecting the spin-splittings
$\sigma^B_{Mn}\omega_{Mn}$ is found to lead to the correct rates only 
for large values of the magnetic field while for smaller magnetic fields
qualitative features, such as the minimum in $T_2^{-1}$, are not obtained.

\begin{figure}[t!]
\includegraphics{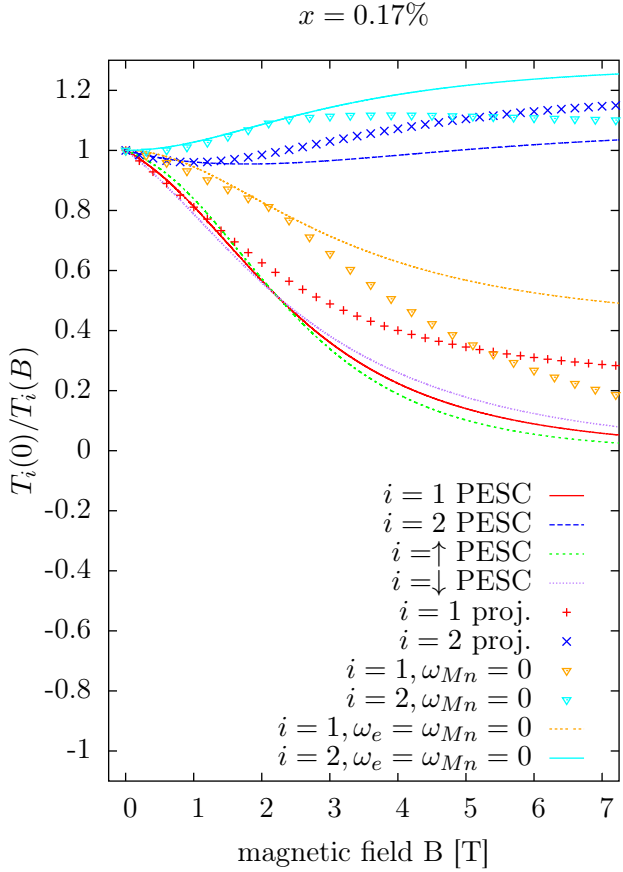}
\caption{Magnetic field dependence of spin transfer rates for
a Cd$_{0.9983}$Mn$_{0.0017}$Te quantum well [cf. Fig.~\ref{fig:sem}]
}
\label{fig:sem_low}
\end{figure}

\begin{figure}[t!]
\includegraphics{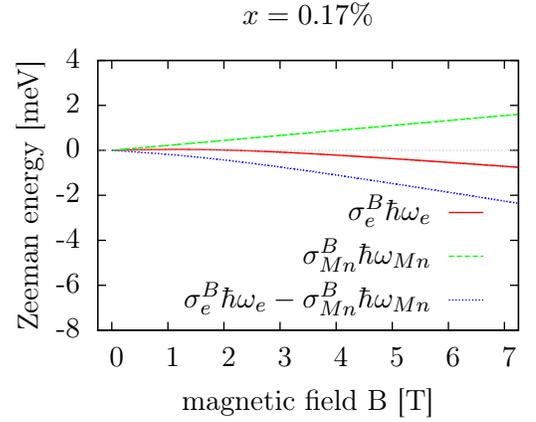}
\caption{Magnetic field dependence of the Zeeman energies for
a Cd$_{0.9983}$Mn$_{0.0017}$Te quantum well [cf. Fig.~\ref{fig:sem_w}]
}
\label{fig:sem_low_w}
\end{figure}

In our analysis of the magnetic field dependence of the spin transfer rates
it was important that $\sigma^B_e\omega_e-\sigma^B_{Mn}\omega_{Mn}>0$.
The situation can change significantly, if this is not the case. In 
order to study this regime of parameters, we repeat the same calculations
shown in Figs.~\ref{fig:sem} and \ref{fig:sem_w} but we assume 
a Mn concentration $x=0.17\%$ which is smaller by a factor of 10 than 
in the previous calculations. The results are displayed in 
Figs.~\ref{fig:sem_low} and \ref{fig:sem_low_w} respectively.
We find in Fig.~\ref{fig:sem_low_w} that now also the electron spin-splitting 
is dominated by the Zeeman term and the mean-field contribution from 
the impurity magnetization is rather small. In particular, one finds that
$\sigma^B_e\omega_e-\sigma^B_{Mn}\omega_{Mn}$ is now negative for all
values of $B>0$.
This fact has immediate consequences on the magnetic field dependence of 
the spin transfer rates. The main qualitative difference between the 
rates shown in Fig.~\ref{fig:sem_low} and in the previous case is that
now the parallel spin transfer rate $T_1^{-1}$ decays to zero for 
large values of $B$. Here, the spin transfer channel corresponding to
$T^{-1}_\up$ is closed due to the energy splitting, whereas
$T^{-1}_\down$ decreases to zero, because the prefactor $\Gamma^+$ tends 
to zero for $B\to \infty$. The physical reason for this behaviour is that
due to the negativity of $\sigma^B_e\omega_e-\sigma^B_{Mn}\omega_{Mn}$
spin-flips from the spin-up to the spin-down band face an energy penalty,
while a flip from the spin-down to the spin-up band would require 
a corresponding decrease of an impurity spin in order to satisfy the 
spin conservation. However, for $B\to \infty$ the impurity spins are
already fully aligned antiparallel to the magnetic field, so that this
spin-flip is also forbidden.
The magnetic field dependence of the perpendicular spin transfer rate
$T_2^{-1}$ for $x=0.17\%$ is quantitatively similar to the case 
of $x=1.7\%$. However, here, the asymptotic value for strong magnetic fields
is $T_2^{-1}(B\to\infty)\to \frac{15}{14}\tau_{FGR}$.

For the smaller impurity concentrations, 
the projection operator method of Ref.~\onlinecite{Semenov} 
overestimates the spin transfer rates.
Fig.~\ref{fig:sem_low} also shows that, in this case, neglecting
the impurity Zeeman terms leads to significant deviations from 
the energy-conserving rates.

In order to establish a connection between the theories discussed above
and the experimentally determined electron spin relaxation rates,
it has to be noted that in most magneto-optical experiments on
II-VI DMS quantum wells so far
(cf. Ref.~\onlinecite{BenCheikh2013} and references therein)
the pump laser is tuned to the electron-heavy-hole exciton energy.
To model these experiments also the Coulomb correlations between 
electrons and holes have to be taken into account, which is beyond 
the scope of the present article. 
It was found in Ref.~\onlinecite{BenCheikh2013} that different groups 
consistently measured perpendicular electron spin relaxation rates $T_2^{-1}$
which are about 5 times larger than $\tau_{FGR}$ at $B=0$.
This discrepancy can be understood
by the fact that the effective electron mass
has to be replaced by the exciton mass in the 
expression for the rate $\tau_{FGR}^{-1}$\cite{BastardExcitonSpinScattTime92},
which yields an increase of the rate by a factor of $\sim4.6$ 
in the case of CdMnTe.
Nevertheless, the finding of the present article that the rate $T_2^{-1}$ 
varies only weakly with the magnetic field and stays essentially 
within $30\%$ of $\tau_{FGR}^{-1}$ is consistent with the tendency of 
most of the experimental results summarized in Ref.~\onlinecite{BenCheikh2013}.
However, especially for samples with low impurity concentration at low
temperatures, there are also some experiments which 
measured a maximum (instead of a minimum predicted
by the present theory) of the magnetic field dependence of the
perpendicular spin transfer rate as well as changes in the rate which span 
about one order of magnitude of its value at $B=0$, which was 
suggested\cite{BenCheikh2013} to stem from local fluctuations of the 
impurity magnetization.
In order to distinguish these imhomogeneity effects from Coulomb correlation 
effects we suggest experiments where the pump pulse is tuned to energies 
well above the exciton resonance.

\subsection{Interplay between $s$-$d$ and Rashba Interactions}
\label{sdSO}
The fact that in the derivation of Eq.~(\ref{eq:QkdepMarkov}) the
$\k$-dependence of an effective magnetic field was taken into account 
makes it possible to discuss the interplay between the spin-orbit coupling
and the $s$-$d$ interaction on a rigorous microscopic basis, where the
spin-orbit interaction also acts during $s$-$d$ scattering events.
In earlier works, the interplay between these effects was 
studied\cite{Ungar15,Proceedings_Pablo_2015}, where only the direct 
effects of the spin-orbit coupling on the electron spins was considered,
yielding an additional $\k$-dependent contribution to the mean-field 
precession frequency, whereas the dynamics of the correlations was not 
modified, i.~e., the spin-orbit interaction was only accounted for between
$s$-$d$ scattering events. It was found that already on a mean-field level, the 
carrier spin dephasing due to the $\k$-dependence of the precession frequencies
can be strongly suppressed by a motional-narrowing-type mechanism caused by
the precession in the mean field of the impurity magnetization. Furthermore,
it was argued that both mechanisms can be tuned in a wide range, 
especially in Hg$_{1-x-y}$Cd$_y$Mn$_x$Te quantum wells with applied electric
fields. In this material, the strength of the $s$-$d$ interaction is 
determined by the Mn concentration $x$, while the Cd concentration $y$ can
be used to change the gap between conduction and valence bands which 
controls the strength of the Rashba\cite{Rashba} field.
When both types of interaction are similarly important, a complex
oscillatory time evolution of the carrier spin was found, which is absent
when either one of the interactions dominates.
\begin{figure}[t!]
\includegraphics{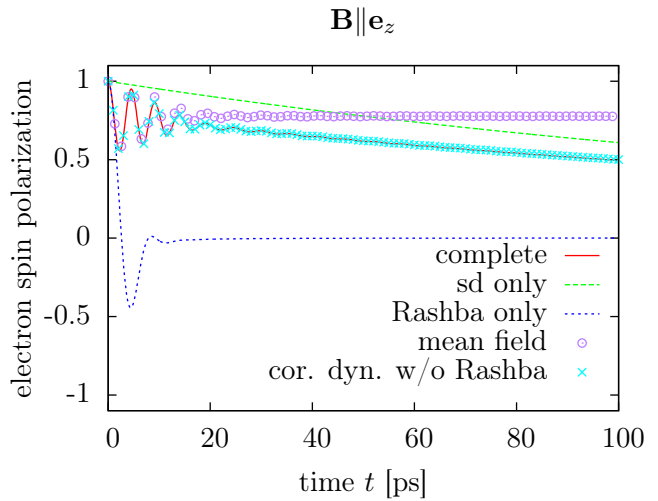}
\caption{
Time evolution of the total electron spin polarization after spin polarized 
optical excitation in a magnetic field perpendicular to the quantum well 
plane (cf. text for parameters). 
The red solid line describes the results according to 
Eqs.~(\ref{eq:eom}b) and (\ref{eq:eom}c) with the Markovian expression 
for the correlations from Eq.~(\ref{eq:QkdepMarkov}). The green dashed line
corresponds to a calculation without Rashba coupling, where only the 
$s$-$d$ interaction is considered. The blue dotted line presents the results
of the case in which only the Rashba interaction is present. The mean-field
calculation, which is obtained by dropping the correlations completely, is
shown as the purple circles. 
The cyan crosses describe the results where
the effects due to the Rashba interaction on the dynamics of the
carrier-impurity correlations are neglected, so that in addition to the
mean-field terms, the time derivative of the carrier variables obtains 
the correlation induced contribution of Eqs.~(\ref{eq:PESC}).
}
\label{fig:kdep}
\end{figure}

Now, the question arises whether neglecting the effects of the Rashba field
on the dynamics of the correlations is indeed a good approximation or if
qualitative changes have to be expected if they are accounted for.
We study this question in a case in which the strengths of the Rashba and
the $s$-$d$ interactions are comparable. We consider a 
$d=20$ nm wide Hg$_{1-x-y}$Cd$_y$Mn$_x$Te quantum well 
with electric and magnetic fields applied along 
the growth direction $z$. The voltage drop between the barriers of the
quantum well leads to a strong Rashba interaction of the form
\begin{align}
&H_R=2\hbar\alpha_R\sum_{\k\sigma\sigma'}\big(k_y s^x_{\sigma\sigma'}
-k_x s^y_{\sigma\sigma'}\big)c^\dagger_{\k\sigma}c_{\k\sigma'},
\end{align}
where we assume a value of $\alpha_R=4.87$ meVnm\cite{Ungar15}.

Further parameters that enter the calculation are the effective mass
$m^*=0.093 m_0$, the $s$-$d$ coupling constant $\Jsd=15$ meV nm$^3$,
the lattice constant $a=0.645$ nm and the Mn concentration 
$x=7\%$. The initial Mn state is modelled by a thermal equilibrium state 
following a Brillouin function with temperature $T=4$ K 
in an external magnetic field pointing in the $-z$-direction
with $|\mbf B|=50$ mT. The g-factors for impurities and conduction band 
electrons are $g_{Mn}=2.$ and $g_e=-1.5$, respectively.
Furthermore, as we consider an intrinsic DMS where 
the quasi-free carriers originate purely from optical excitation, 
$\NMn\gg N_e$ is clearly fulfilled, so that we can neglect the back-action 
of the carriers on the impurities. Thus, the Mn magnetization remains 
homogeneous, which allows us to integrate along the growth direction
yielding a factor of $I=1.5$. The initial electron spin was modelled by
a Gaussian distribution in the spin-up band centered at the band edge
with standard deviation $E_s=1$ meV corresponding to a $\sigma^-$ polarized
laser with pulse duration (FWHM)  $\sim 140$ fs.
For these parameters, the mean-field energy splitting caused by 
the impurity magnetization is $\sim -0.75$ meV 
(the spin-up-subband is energetically favored), while the strength of
the Rashba interaction for an electron with kinetic energy
$\frac{\hbar^2 k_0^2}{2m^*}=1$ meV is $2\hbar\alpha_R k_0\sim 0.89$ meV.
Here, the Zeeman terms yield significantly smaller contributions
of $g_e \mu_B |\mbf B|\approx -0.004$ meV and 
$g_{Mn} \mu_B |\mbf B|\approx 0.006$ meV to the respective spin splittings.
\begin{figure}[t!]
\includegraphics{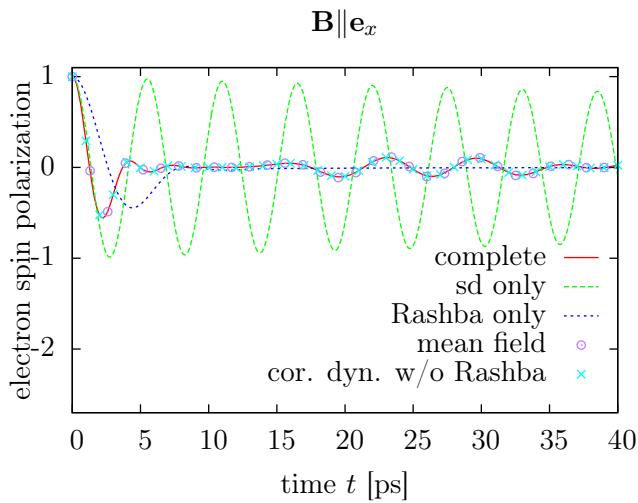}
\caption{Time evolution of the total electron spin polarization after 
spin polarized optical excitation in a magnetic field parallel to the 
quantum-well plane (cf. Fig.~\ref{fig:kdep})
}
\label{fig:kdep_x}
\end{figure}

Fig.~\ref{fig:kdep} shows the results of numerical simulations for this set
of parameters. As reported earlier\cite{Ungar15}, the Rashba interaction 
alone (blue dashed line) leads to a fast dephasing of the carrier spins.
If additionally magnetic impurities with a finite magnetization are present,
already a mean-field treatment (purple circles) can lead to a
strong suppression of the dephasing by motional narrowing caused by the 
precession of the carrier spin in the mean field of the impurity magnetization.
Without the Rashba interaction, the $s$-$d$ interaction causes a spin transfer
from the carriers to the impurities which can be seen in Fig.~\ref{fig:kdep}
as an exponential decay to a non-vanishing equilibrium value. 
In the previous studies\cite{Ungar15}, 
the correlation induced spin transfer was combined
with the mean-field precession, but the effects of the Rashba interaction on
the dynamics of the correlations were neglected (here shown as cyan crosses). 
In Fig.~\ref{fig:kdep},
also the complete carrier spin dynamics is shown, where the Rashba interaction
is explicitly accounted for in the calculation of the correlations (red solid
line).
By comparing both calculations, it can be seen that the total carrier spin
is hardly influenced by the effects of the Rashba spin-orbit coupling on
the correlation dynamics. The same result is also obtained for the 
situation where the magnetic field is applied parallel to the quantum
well plane, as shown in Fig.~\ref{fig:kdep_x}.

\begin{figure}[t!]
\includegraphics{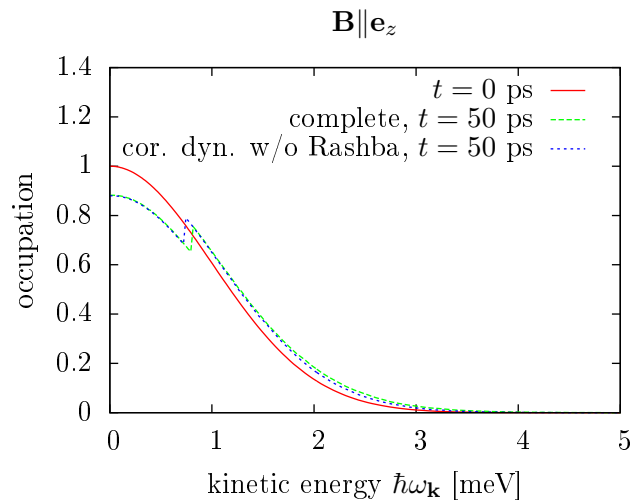}
\caption{Kinetic energy dependence of the occupation of carrier states
at times $t=0$ ps and $t=50$ ps for the calculations shown in 
Fig.~\ref{fig:kdep}.
}
\label{fig:kdep_occup}
\end{figure}

Similar to the fact that the precession-type motion of the correlations 
discussed so far leads to changes in the kinetic energy of scattered carriers, 
also the Rashba interaction enforces a precession of the
correlations resulting in corresponding changes in the electron energies.
In Fig.~\ref{fig:kdep_occup} the carrier occupations at $t=0$ and $t=50$ ps 
are shown for calculations with and without accounting for the
Rashba effect on the correlation dynamics for the situation described in
Fig.~\ref{fig:kdep} with magnetic field parallel to the growth direction.
Without the Rashba interaction, the kinetic energy dependence of the
occupations at $t=50$ ps shows a distinctive step at 
$\hbar\omega_{\k}=|\hbar\sigma^B_e\omega_e-\hbar\sigma^B_{Mn}\omega_{Mn}|$ 
which corresponds to a redistribution of carriers with an excess 
energy in the spin degree of freedom to states with higher kinetic energies.
When the Rashba coupling is turned on, the step shifts towards slightly 
higher kinetic energies. This can be explained by the fact that
in the configuration with a magnetic field along the growth direction and
a Rashba field in the quantum well plane the energy eigenvalues of an 
electron with wave vector $\k$ are
\begin{align}
E_\pm=\hbar\omega_\k\pm\frac 12 \hbar\sqrt{\big(2\alpha_R |\k|\big)^2+
\big(\sigma^B_e\omega_e\big)^2}.
\end{align}
Including the shifts due to the impurity Zeeman splittings, 
the step in the kinetic energy dependence
of the occupation is therefore shifted to
$\hbar\sqrt{\big(2\alpha_R |\k|\big)^2+
\big(\sigma^B_e\omega_e-\sigma^B_{Mn}\omega_{Mn}\big)^2}$.
However, the shift of the energy splitting is too small to cause 
a significant impact on the time evolution of the total spin.

\subsection{Connection to the theory of collective carrier-impurity 
precession modes in DMS}
In the derivation of the theory, the $z$-dependence of the carrier envelope
function was taken into account. We see from Eqs.~(\ref{eq:rate_expl}) that
one effect of this $z$-dependence is that the spin transfer rate obtains
the prefactor $I$.  Assuming a constant linear impurity density
$\frac{\NMn}d$, a constant $z$-envelope yields a value of $I=1$ while
the extreme case of a quantum well with infinite barriers yields
$I=\frac 32$. This effect has also been found in previous 
studies of DMS~\cite{CdMnTeFerroMag97,Semenov,BenCheikh2013}.

Like the spin transfer rates, also the electron spin precession
is influenced by the $z$-dependence of the envelope of the electron wave
function. In particular, it can be seen from Eq.~(\ref{eq:def_om_e}) that 
the contribution to the electron spin precession frequency from the
impurity spin is proportional to 
$\int\limits_{-\frac d2}^{\frac d2}dz\, |\psi(z)|^2 \langle \mbf S(z)\rangle$. 
Thus, the impurity spin as a function of $z$ can
be decomposed into this mode which couples to the electron spin precession
and $\NMn-1$ orthogonal modes, which do not influence the electron spins
directly on a mean-field level.
In the parameter regime where the precession frequencies of the electron 
and impurity spins almost coincide, the coupling between the above impurity
mode and the electron spin is particularly large, leading to an
avoided crossing indicating a collective motion of impurity and carrier
spins. This fact has been discussed in a number of recent articles
by different groups\cite{Teran03, ColMod2,SpinWavesPerezCibert11,
Shmakov11,Barate10,Frustaglia04}. In these works, however, the 
carrier-impurity correlations have been disregarded.

In the following, we will derive equations describing the situation studied,
e.g., in Ref.~\onlinecite{ColMod2} taking the effects due to 
the correlations into account.
There, a n-type \mbox{CdMnTe} quantum well in an external
magnetic field parallel to the quantum well plane ($x$-direction) 
was considered, leading to equilibrium 
values of the impurity and carrier spins antiparallel to the magnetic field. 
A circularly polarized pump beam induces electron-hole pairs with
spin polarization along the $z$-direction. 
During the decay of the hole spins on a timescale of $\sim 5$ ps, 
the impurity magnetization precesses around the $p$-$d$ exchange field 
of the holes, causing a small tilt of the impurity spins away from the
equilibrium $x$-axis into the $y$-axis. The optically induced electron spins
contribute to the $z$-component of the total carrier spin. Thus, after the
holes are decayed, one ends up with a situation where the impurity and 
carrier spins precess around each other.

The fact that the spin components perpendicular to the equilibrium values
are small compared with the parallel components allows one to linearize
Eqs.~(\ref{eq:PESC}) and (\ref{eq:PESCMF}) with the expression for the rates
from Eq.~(\ref{eq:rate_expl}):
\begin{widetext}
\begin{subequations}
\begin{align}
\ddt \mbf s^\perp_{>/<}&=
\frac{g_e\mu_B}\hbar\mbf B \times \mbf s^\perp_{>/<} 
+\frac{\Jsd\NMn}{\hbar V}  \mbf S^{x,(1)}\times \mbf s^\perp_{>/<}
-\frac{\Jsd\NMn}{\hbar V} \mbf s^x_{>/<} \times \mbf S^{\perp,(1)}
-\frac 1d\int\limits_{-\frac d2}^{\frac d2}dz\, \Gamma^{>/<}(z)
 \mbf s^\perp_{>/<}\\
\ddt \mbf S^{\perp,(j)}&= 
\frac{g_{Mn}\mu_B}{\hbar}\mbf B\times \mbf S^{\perp,(j)} 
-\frac{\Jsd}{V\hbar} \mbf S^{x,(j+1)}\times (\mbf s^\perp_{>}
+\mbf s^\perp_<)
+\frac{\Jsd}{V\hbar} (\mbf s^x_{>}+\mbf s^x_{<})\times \mbf S^{\perp,(j+1)}
+\nn&
+d^{j-1}\int\limits_{-\frac d2}^{\frac d2}dz\, |\psi(z)|^{2j} 
\big(\Gamma^>(z) \mbf s^\perp_>+\Gamma^<(z) \mbf s^\perp_<\big)
\\
\Gamma(\k,z):&=
\pi \frac{A m^*}{2\pi\hbar} \frac{\Jsd^2\NMn}{\hbar^2 V^2} 
d^2|\psi(z)|^4 \bigg[
\langle {S^\|}^2\rangle +
\Big(\frac{\langle{S^\perp}^2\rangle}2
-\sigma^B_S\frac{|\langle\mbf S\rangle|}4\Big)
\Theta\big(\omega_\k+(\sigma^B_e\omega_e-\sigma^B_e\omega_{Mn}(z))\big) +
\nn&+
\Big(\frac{\langle{S^\perp}^2\rangle}2
+\sigma^B_S\frac{|\langle\mbf S\rangle|}4\Big)
\Theta\big(\omega_\k-(\sigma^B_e\omega_e-\sigma^B_e\omega_{Mn}(z))\big)\bigg]
\end{align}
\label{eq:sw_rewr}
\end{subequations}
\end{widetext}
with
\begin{subequations}
\begin{align}
\mbf s^{x/\perp}_{>/<}:&={\sum_{\k}}^{>/<}  \mbf s^{x/\perp}_\k \\
\Gamma^{>/<}(z):&={\sum_{\k}}^{>/<} \Gamma(\k,z)  \\
\mbf S^{x/\perp,(j)}:&=d^{j-1}\int\limits_{-\frac d2}^{\frac d2} dz\,
|\psi(z)|^{2j} \langle \mbf S^{x/\perp}(z)\rangle
\end{align}
\end{subequations}
where the indices $x$ and $\perp$ denote the spin components parallel 
and perpendicular to the equilibrium axis $x$ and ${\sum_{\k}}^{>/<}$
describes the sum over all wave vectors $\k$ with $\omega_\k>\omega_0$
or $\omega_\k<\omega_0$, respectively, where 
$\omega_0=|\sigma^B_e\omega_e-\sigma^B_{Mn}\omega_{Mn}|$.
The distinction between states with higher or lower kinetic energy than
$\omega_0$ is a direct consequence of the step-like $\k$ dependence of the 
spin transfer rates.

Eqs.~(\ref{eq:sw_rewr}) of the present paper differ mainly from 
Eqs.~(4) and (5) of Ref.~\onlinecite{ColMod2} in that 
carriers with $\omega_\k <\omega_0$ are distinguished from 
carriers with $\omega_\k >\omega_0$ and in that
the terms proportional to the rate $\Gamma^{>/<}(z)$ are omitted
in the mean-field treatment of Ref.~\onlinecite{ColMod2}. Instead, a 
phenomenological relaxation rate $\tau_e^{-1}$ was added manually
in Ref.~\onlinecite{ColMod2}.
Another difference is the appearance of the corresponding spin transfer
term in the equations for the impurities. This is due to the fact that
the $s$-$d$ interaction is spin conserving so that the electron spin that
is removed from $\mbf s^\perp_{>/<}$ has to be transferred to the impurity
system. 
Taking these corrections with respect to the description of 
Ref.~\onlinecite{ColMod2} into account would lead to a more accurate 
modelling of the collective carrier-impurity precession modes.
However, as discussed earlier, the variation of the 
perpendicular spin transfer rate in the presence of an external magnetic
field is limited to $\lesssim 30\%$ of the golden rule value at $\mbf B=0$, so 
that the spin transfer rate remains in the same order of magnitude. Thus,
the phenomenological treatment of the rate 
can be justified for the purpose of the discussion in
Ref.~\onlinecite{ColMod2}.

\section{Conclusion}
A quantum kinetic description of the carrier spin dynamics in 
paramagnetic intrinsic II-VI DMS was presented which, in contrast to
previous works\cite{Thurn:12,Cygorek:14_1,PESC}, 
also accounts for a wave-vector dependent effective magnetic field 
as well as Zeeman terms for carriers and impurities.
The Markov limit of the quantum kinetic equations allow us to 
extract rates for spin transfer processes between carriers and magnetic
impurities. 
From the rigorous treatment of a precession-type dynamics of the 
carrier-impurity correlations
it is found that the redistribution of carriers in $\k$-space is not only 
influenced by the spin-splitting of the electron subbands due to the 
Zeeman energy enhanced by the impurity magnetization, but also acquires an
energetic shift corresponding to the Zeeman level splitting of the 
magnetic impurities. 
This shift accounts for the fact that a spin flip of an electron
involves a spin flop of the magnetic impurity in the opposite direction
and the total energy of the magnetic impurity and the electon spin
has to be conserved.
The energetic shifts in the description of the spin flip-flop processes 
are often not correctly accounted for in the literature. 

The impact of these energy shifts was investigated using the example of the
magnetic-field dependence of the carrier-impurity spin transfer rates
parallel and perpendicular to the impurity magnetization. Two distinct 
parameter regimes were identified, one for rather high doping concentrations 
of the order of $x\sim 1\%$ and one for extremely diluted systems 
with $x\lesssim 0.1\%$.
These regimes correspond to cases where
the total change of the kinetic electron energy 
as given by $(\sigma^B_e\omega_e-\sigma^B_{Mn}\omega_{Mn})$ 
is mainly positive or negative. 
In both situations the perpendicular spin transfer rate 
$T_2^{-1}$ varies within $\sim 30\%$ of the value for $B=0$, which 
also coincides with the results for $T_1^{-1}$ 
obtained by Fermi's golden rule.
However, in the first case, the parallel spin transfer 
rate $T_1^{-1}$ decays monotonically for an increasing magnetic 
field to $\frac 58$ of the Golden Rule value at $B=0$, while in the extremely
diluted case, $T_1^{-1}$ eventually vanishes. 
In calculations where the carrier spin splitting $\hbar\omega_e$ 
or the impurity Zeeman splitting $\hbar\omega_{Mn}$ is neglected, 
as is often done in the literature, 
the magnetic-field dependence of the spin transfer rates 
deviates significantly from that predicted by the accurate description
involving both energetic shifts.
Accounting for the impurity Zeeman splitting for the spin flip-flop 
processes turns out to be particularly important in the
very dilute case.

Furthermore, the interplay between the $s$-$d$ interaction between carrier and
impurites and the Rashba interaction in a 
Hg$_{1-x-y}$Cd$_y$Mn$_x$Te quantum well was investigated.
In the standard rate description approach one usually calculates for each  interaction
a corresponding scattering rate and ignores that other interactions might change the
dynamics {\em during} the scattering process. This was the point of view adopted in previous studies
of the combined dynamics of $s$-$d$ and Rashba couplings\cite{Ungar15,Proceedings_Pablo_2015}.
However, such mutual dependencies of different interactions have been shown in the literature 
to be of importance, e.g., in the case of a static electric field acting during phonon scattering process
 known as {\em intracollisional field effects} \cite{RossiICFE}.
Technically, the dynamics during an ongoing interaction process is represented by correlation functions.
In the present article, we presented a quantum kinetic description where 
$s$-$d$ and Rashba interactions have been fully accounted for in the combined dynamics of the single-particle 
density matrices and the carrier-impurity correlations, thus fully covering all mutual cross-effects between
these interactions. While it is {\em a priori} difficult to predict how important these cross-effects actually are,
we have demonstrated for the present case
that the total carrier spin is hardly affected by this mechanism.

Finally, taking into account also the $z$-dependence of the carrier
envelope function makes it possible to show how the 
phenomenological treatment of the spin transfer rate in the description of 
collective carrier-impurity precession modes in Ref.~\onlinecite{ColMod2}
can be based on a solid microscopic foundation.

In summary, our microscopic treatment  
of the effects of a $\k$-dependent magnetic field and 
the impact of the shape of the carrier envelope function 
justifies the approximations made in earlier studies
of the dynamics of the total electron spin\cite{Ungar15,ColMod2}.
Apart from this new insight,  the present theory further contributes to 
the progress in the field of  spin physics in DMS by
not only deriving rates for carrier spins parallel, but also perpendicular to
the impurity magnetization in the presence of an external magnetic field.
The latter are expected to be the dominant contribution to the 
carrier dephasing time in time-resolved magneto-optical Kerr measurements
in Voigt configuration. 
In contrast to earlier approaches found in the literature\cite{Semenov,Wu09}, 
the rates derived in this article are fully compatible with
the energy conservation of an individual spin flip-flop process.
Our study reveals that the difference between the predictions of the discussed theories 
is most prominent for extremely diluted magnetic semiconductors.

\begin{acknowledgments}
We gratefully acknowledge the financial support from the Universidad de
Buenos Aires, project UBACyT 2014-2017 No. 20020130100514BA, 
and from CONICET, project PIP 11220110100091.
\end{acknowledgments}

\bibliography{alle}
\end{document}